\documentclass[prx,reprint,superscriptaddress,longbibliography]{revtex4-2}

\usepackage{hyperref}       
 \hypersetup{
    colorlinks=true,
    linkcolor=magenta,
    filecolor=magenta,      
    urlcolor=blue,
    citecolor=red,
    pdftitle={Generation and structuring of multipartite entanglement in Josephson parametric system},
    pdfpagemode=FullScreen,
    }
\usepackage{url}            
\usepackage{amsmath,amssymb}
\usepackage{booktabs}       
\usepackage{amsfonts}       
\usepackage{graphicx}
\usepackage[mathscr]{euscript}
\usepackage[caption=false]{subfig}

\usepackage[export]{adjustbox}
\usepackage{xcolor}
\graphicspath{ {./fig/} } 

\usepackage[fleqn,tbtags]{mathtools}
\usepackage{amsbsy}

\begin{document}

	\title{Generation and structuring of multipartite entanglement in\\ a Josephson parametric system
	}
	
\author{K. V. Petrovnin}%
 \email[Correspondence: ] {kirill.petrovnin@aalto.fi}
 \affiliation{
  QTF Centre of Excellence, Department of Applied Physics, Aalto University, P.O. Box 15100, FI-00076 AALTO, Finland
  }

\author{M. R. Perelshtein}
    \affiliation{
    QTF Centre of Excellence, Department of Applied Physics, Aalto University, P.O. Box 15100, FI-00076 AALTO, Finland
    }

\author{T. Korkalainen}%
 \affiliation{
  QTF Centre of Excellence, Department of Applied Physics, Aalto University, P.O. Box 15100, FI-00076 AALTO, Finland
  }
  
  \author{V. Vesterinen}%
 \affiliation{QTF Centre of Excellence, VTT Technical Research Centre of Finland Ltd,\\ P.O. Box 1000, FI-02044 VTT, Finland
 }

\author{I. Lilja}%
 \affiliation{
  QTF Centre of Excellence, Department of Applied Physics, Aalto University, P.O. Box 15100, FI-00076 AALTO, Finland
  }
  
\author{G. S. Paraoanu}%
 \affiliation{
  QTF Centre of Excellence, Department of Applied Physics, Aalto University, P.O. Box 15100, FI-00076 AALTO, Finland
  }
  \affiliation{InstituteQ – the Finnish Quantum Institute, Aalto University, Finland}
  
 \author{P. J. Hakonen}%
 \email[Correspondence: ] {pertti.hakonen@aalto.fi}
 \affiliation{
  QTF Centre of Excellence, Department of Applied Physics, Aalto University, P.O. Box 15100, FI-00076 AALTO, Finland
  }
  \affiliation{InstituteQ – the Finnish Quantum Institute, Aalto University, Finland}
  
\begin{abstract}
	{
		Quantum correlations are a vital resource in advanced information processing based on quantum phenomena. 
		Remarkably, the vacuum state of a quantum field may act as a key element for the generation of multipartite quantum entanglement. 
		In this work, we achieve generation of genuine tripartite entangled state and its control by the use of the phase difference between two continuous pump tones. We demonstrate control of the subspaces of the covariance matrix for tripartite bisqueezed state.
		Furthermore, by optimizing the phase relationships in a three-tone pumping scheme we explore genuine quadripartite entanglement of a \textit{generalized} H-graph state ($\mathscr{\tilde{H}}$-graph). 
		Our scheme provides a comprehensive control toolbox for the entanglement structure and allows us to demonstrate, for first time to our knowledge, genuine quadripartite entanglement of microwave modes.
		All experimental results are verified with numerical simulations of the nonlinear quantum Langevin equation. 
		We envision that quantum resources facilitated by multi-pump configurations offer enhanced prospects for quantum data processing using parametric microwave cavities.
	}

\end{abstract}

\maketitle

\section{Introduction}

Entanglement is a crucial resource in advanced information processing based on quantum mechanical concepts \textcolor{blue}{\cite{jozsa2003role,RevModPhys.81.865}}. 
To exceed the computational power of classical devices, quantum devices typically need to employ highly entangled states.  
Controlled generation of entanglement is the key resource not only in quantum computing \textcolor{blue}{\cite{PhysRevLett.86.5188,menicucci2008one,ladd2010quantum,Menicucci2014,pfister2019continuous,40yearsQC}}, but also in sensing \textcolor{blue}{\cite{Giovannetti2011,PhysRevLett.114.110506,RevModPhys.89.035002,RevModPhys.90.035005}} and secure communications \textcolor{blue}{\cite{PhysRevA.74.062305,Gisin2007,yin2020entanglement}}.

One of the most easily accessible and, at the same time, reliable sources for entanglement generation is the vacuum state of a quantum field \cite{SUMMERS1985257}.
Squeezing, which is the fundamental operation for the continuous variable (CV) states production, allows generation of coherence and entanglement from vacuum fluctuations \cite{braunstein2005quantum}.
While two-mode squeezing produces bipartite quantum states, multipartite states can be generated by applying similar operations \cite{lahteenmaki2016coherence}.
Intriguingly, multipartite CV states are shown to enable various promising phenomena such as quantum state sharing \cite{PhysRevA.71.033814} and secret sharing \cite{PhysRevA.59.1829}, dense coding \cite{PhysRevA.61.042302}, error correction \cite{PhysRevA.97.032335} and quantum teleportation \cite{PhysRevA.60.937}. Alongside with phase sensing \cite{guo2020distributed} and quantum sensor networks \cite{PhysRevA.98.012114}, multipartite entangled states have significant potential in multiparameter quantum metrology applications \cite{PhysRevA.98.012114,gessner2020multiparameter}.
Furthermore, CV cluster states show potential as a universal quantum computing platform \cite{Zhang2006,Menicucci2006,Gu2009}, which has been under active development for the past 20 years. 
Cluster state calculus, foremost utilizing optical resources \textcolor{blue}{\cite{mantri2017universality,Menicucci2006,walther2005experimental,PhysRevA.102.052424,PhysRevLett.126.020402}}, realize measurement-based quantum computing \cite{PhysRevLett.86.5188}.

While optical-mode schemes for generation of multipartite states lack versatility, in-situ tunability and are limited to optical frequencies, the microwave platform allows for full control of operations via input rf-signals and integration with the existing silicon-based circuitry. 
During the past few years, significant progress in processing of CV multipartite states at microwaves has been achieved;
for instance, squeezed states produced by microwave cavities have been shown to exhibit correlations between photons in separate frequency bands \cite{lahteenmaki2016coherence} and strong entanglement between different modes \cite{Wilson2018generatingmultimode}.

In this work, we experimentally generate genuinely entangled tripartite and quadripartite  states using a superconducting parametric cavity, operated under steady-state conditions. 
Using the Gaussian-mode formalism \cite{Adesso2007,Weedbrook2012,Fabre2020}, we characterize the generated states and verify entanglement from the covariance matrix.
We develop an analytical description, which allows us to determine the entanglement structure of the generated state and establish a connection to H(amiltonian)-graph representation \cite{Gu2009,menicucci2011graphical,bruschi2016towards,pfister2019continuous}.
All of the experimental results are in good agreement with the theoretical predictions, in which all circuit parameters were set in accordance with the measured characteristics of the device.

The paper is organized as follows.
Section \ref{theorFound} describes the quantum dynamics of a Josephson parametric system and demonstrates the generation protocol for CV multipartite states, such as fully inseparable and genuinely entangled states. Using analytical methods, we provide the entanglement structure, which is described by graphs and their corresponding adjacency matrices.
In section \ref{Experiment} we present the experimental setup and explain our data analysis methods.
In Section \ref{entanglement_results} we present the experimental and theoretical results on the generation of multipartite entanglement using a Josephson parametric system in both tripartite and quadripartite cases.
In Appendices \ref{methods}A-E we present details of our analysis, experimental techniques, and parameter extraction. Appendix \ref{methods}F deals with analytical solution of the equations of motion for tri-/quadripartite states using simplified linear model of parametric amplifier and establish correlations (graph connections) between spectral modes, which allows us to classify and control the entanglement structure. Furthermore, analytical forms of relevant covariance matrices are given.

\section{Theoretical foundations}\label{theorFound}

\subsection{Quantum dynamics of the device}
Our work employs a parametric system comprised of a superconducting $\lambda/4$ resonator terminated in a SQUID loop. Such a setup forms the archetype of a narrow-band superconducting Josephson parametric amplifier (JPA) \cite{yamamoto2016parametric,Wang2019quantum_dynamics}.
In our setting, we pump the JPA using multi-tone external RF magnetic flux through the SQUID at frequencies that are approximately twice the frequency of the resonator $\omega_d \sim 2 \omega_r$ (three-wave mixing) \cite{lahteenmaki2016coherence,Korkalainen2021}. 
In the rotating frame, the Hamiltonian of the system, as derived in Refs.\,\onlinecite{lahteenmaki2013dynamical,yamamoto2016parametric}, is given by:
\begin{eqnarray}
	{H}_{\text{sys,rwa}}(t)={\hbar}\Delta_r \tilde{a}^\dagger \tilde{a} & +\nonumber \\ \frac{\hbar}{2}\sum_{d=1}^{p}(\alpha^{*}_d e^{i\Delta_d t}\tilde{a}^{2} + \alpha_d e^{-i\Delta_d t} \tilde{a}^{\dagger 2}) + 6{\hbar}{K} \tilde{a}^{\dagger}\tilde{a}^{\dagger} \tilde{a} \tilde{a},
\end{eqnarray}
where $\tilde{a}$ ($\tilde{a}^{\dagger}$) is the annihilation (creation) operator for cavity photons in the rotating frame at angular frequency $\omega_{\Sigma}/2$, $\alpha_d$ is the  complex amplitude  of the $d$-th pump tone and $\Delta_d=\omega_d-\omega_{\Sigma}$ is the angular frequency detuning of the corresponding tone.
Possible extra phase factors in different pump tones are included in the complex pump amplitude $\alpha_d=|\alpha| e^{i\varphi_d}$. Here $\Delta_r$ denotes the detuning between half of the average pump angular frequency $\omega_\Sigma$ and the resonator angular frequency $\omega_r$: $\Delta_r=\omega_r-\omega_{\Sigma}/2$, with $\omega_{\Sigma}$ representing the average angular frequency in multi-tone driven case: $\omega_{\Sigma}=(1/p)\sum_{d=1}^{p}\omega_d$ with $d=\left\lbrace  1,\dots,p \right\rbrace  $ as the pump tone index.

Strongly driven SQUIDs are notoriously nonlinear. Therefore, we also include the nonlinear Kerr term with strength $K$ to the description of our parametric system.
The Kerr constant controls the parametric behavior close to and above the critical pumping threshold $\alpha \geq \alpha_{crit}$.
Several effects are accounted for by the Kerr nonlinearity, such as limited maximum gain, compression, observed at $\alpha\lesssim\alpha_{crit}$, broadening and shifting of the resonance curve, 
and parametric oscillation above the critical point.
In our experiment, we employ the pump-power-dependent gain coefficient to extract the Kerr constant (see Appendix \ref{systpara}).

In order to describe the coupling of the cavity resonator to the incoming transmission line and to an intrinsic thermal bath, we include two additional terms in the full Hamiltonian:
\begin{equation}
	{H}(t)=H_{\text{sys,rwa}}(t)+H_{\text{sig}}+H_{\text{loss}},
\end{equation}
where $H_{\text{sig}}$ includes the coupling to the signal port transmission line with dissipation rate $\kappa$, while $H_{\text{loss}}$ includes the coupling to the internal loss port with linear dissipation rate $\gamma$.

Using the Quantum Langevin Equation (QLE), we obtain the output modes in our parametric system.
We employ the standard input/output formalism in the rotating frame, which yields
\begin{eqnarray}
	\label{QLE-NONLINEAR}
\dot{\tilde{a}}(t)=(-i\Delta_r &- \frac{\kappa+\gamma}{2})\tilde{a} - i \sum_{d=1}^{p}\alpha_d e^{i\Delta_d t} \tilde{a}^\dagger \nonumber \\
& + \sqrt{\kappa}\tilde{b}_{in}+\sqrt{\gamma}\tilde{c}_{in} - 12i{K} \tilde{a}^{\dagger} \tilde{a} \tilde{a} 
\end{eqnarray}
where $\tilde{b}_{in}$ and $\tilde{c}_{in}$ are the ladder (annihilation) operators for the signal and linear dissipation ports, respectively. 

The output mode $\tilde{b}_{out}$ is obtained using the following relation between the incoming and outgoing modes:
\begin{equation}
\label{IOeq}
	\tilde{b}_{out}(t)=\tilde{b}_{in}(t)-\sqrt{\kappa}\tilde{a}(t).
\end{equation}
We are interested in the correlations embedded in the output mode given by Eq. (\ref{IOeq}) in the time domain. The correlations 
can be revealed in full after Fourier transformation to the frequency domain.
By defining finite-band spectral modes (see Section \ref{spectral_mode_def}) in the frequency domain and examining correlations between these spectral ranges, we can verify the presence of entanglement in the band-limited microwave signals.
\begin{figure}
   \includegraphics[height=6.5cm]{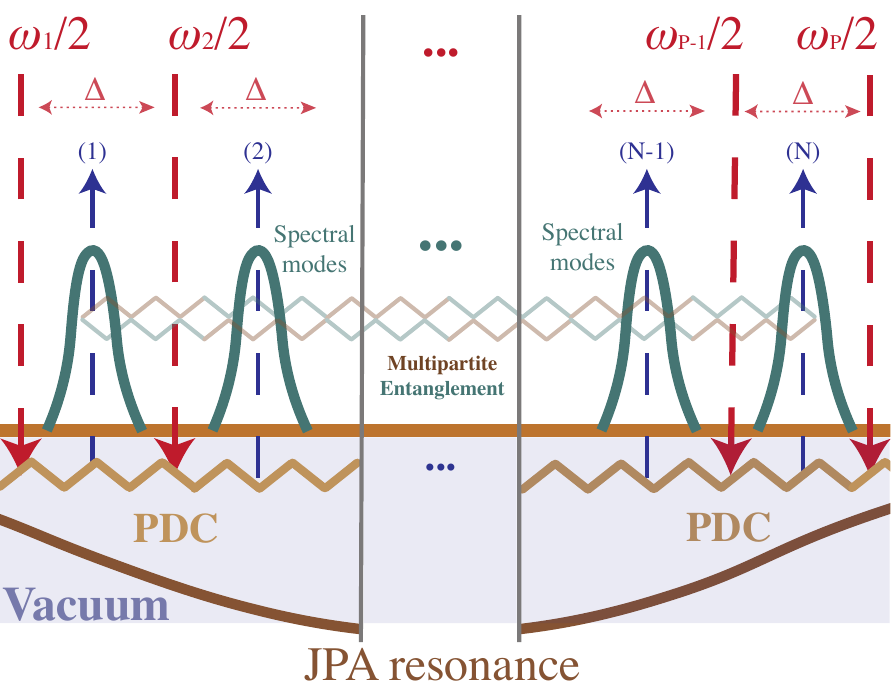}
	\caption{
    {\bf Definition of spectral modes and their correlations in a multi-tone pump setting.}
    Spectral modes in a multi-pump JPA, where the pump tones (red arrows) trigger parametric downconversion process (PDC) leading to the appearance of multipartite correlation between microwaves (blue arrows), extracted from vacuum fluctuations.
    Numbered spectral modes depicted in green are also correlated due to the continuous pumping of the JPA resulting in  multipartite entanglement between microwaves in the spectral modes. The bandwidth of each spectral mode $\Delta$ in Fourier analysis is chosen to be much narrower than the cavity resonance width.
}
	\label{modes_separation_fig}
\end{figure}

\subsection{Spectral modes definition} \label{spectral_mode_def}

Parametric downconversion processes and the definition of employed spectral modes within the fundamental cavity resonance in a multi-pump JPA are illustrated in Fig.\,\ref{modes_separation_fig}. 
The spacing and width of the spectral modes are selected in such a manner that that the modes are generated within the linewidth of the JPA resonance.
Each pump that acts on the JPA triggers spontaneous parametric downconversion of a pump photon (vertical red arrows) into two photons, with their energies summing up to the energy of the pump photon (blue arrows). 
This process is stimulated by vacuum fluctuations, whose existence is a fundamental feature of quantum electrodynamics.
One might expect that the downconversion processes would be random, occurring independently for each pump in the multi-tone pumping situation. 
In such a case, the result would simply be a sum of the downconversion processes, but this turns out not to be the case.
Instead, the photons are fundamentally correlated, even if they originate from different pump tones, because they were "born into existence" by the same quantum fluctuation.
In other words, one spectral mode contains photons correlated with the quanta in the other spectral modes and, consequently, we expect multipartite correlations to appear (depicted schematically via the zigzag lines).

We consider the fundamental resonance of a transmission line JPA centered at $\omega_r$ frequency with a bandwidth $2\delta\omega$, within which ($[\omega_r-\delta\omega, \omega_r+\delta\omega]$) we define $N$ spectral modes as depicted in Fig.\,\ref{modes_separation_fig}. 
Let us define $\tilde{a}$ as a vector of spectral modes:
\begin{equation}
\label{mode_expansion}
    \tilde{a}=\lbrace \tilde{a}_1,\dots,\tilde{a}_N,\tilde{a}^\dagger_1,\dots,\tilde{a}^\dagger_N\rbrace^T,
\end{equation} 
where $N$ is a total mode number and the creation $\tilde{a}_i^{\dagger}=\tilde{a}_i^{\dagger}(t)$ and annihilation $\tilde{a}_i=\tilde{a}_i(t)$ operators are time-dependent.
In general, the frequency difference between half of $p$-th and $(p+1)$-th pump tones defines the bandwidth of the spectral mode $\tilde{a}_i$. 
We employ an equidistant pump scheme where bandwidth of each mode $\tilde{a}_i$ is defined as $\Delta$ such that the spectrum of a full set of modes $\tilde{a}_i$ covers the bandwidth $2\delta\omega$ of the cavity mode $\tilde{a}$.
In the experiment, we collect the emitted power over the whole $[\omega_r-\delta\omega, \omega_r+\delta\omega]$ frequency range and separate signals in the $N$ modes using numerical postprocessing.
The same operation could be implemented by using accurate bandpass filters with bandwidth $\Delta$.

Within the scope of this work, we consider only tripartite ($N=3$, $p=2$) and quadripartite ($N=4$, $p=3$) quantum states.
In the following, we elucidate the internal structure of the generated states through a graph representation based on the quantum Langevin equation.

\subsection{Graphical description of quantum states}

In order to construct a comprehensive graphical representation, we consider interactions between cavity and input vacuum modes by solving the QLE given in Eq.\,\eqref{QLE-NONLINEAR} for $N$ cavity modes defined in Eq.\,\eqref{mode_expansion}; for details see Appendix \ref{analytics}.
Assuming the strong coupling regime with $\kappa \gg \Delta$ while neglecting dissipation losses $\gamma$ and the Kerr-nonlinearity $K$, Fourier transformation yields a system of linear equations which can be cast in a matrix form:
\begin{equation}
\label{M}
    \mathbf{M}\tilde{a}(\omega)=\sqrt{\kappa}\tilde{b}_{\text{in}}(\omega).
\end{equation}
Here the interaction matrix $\mathbf{M}$ contains diagonal entries provided by cavity-related part of the Hamiltonian, whereas the parametric terms appear in the off-diagonal entries. 
For example, one obtains for the tripartite case with $\Delta_r=0$ and having different phases for the pump tones $\alpha_1=\alpha e^{i\varphi_1}, \alpha_2 =\alpha e^{i\varphi_2}$:
\small
\begin{equation}
\mathbf{M}=
\resizebox{.82\linewidth}{!}{%
  $  \begin{bmatrix}
c & 0 & 0 & 0 & i\alpha e^{i\varphi_1} & 0\\
0 & c & 0 & i\alpha e^{i\varphi_1}  & 0 & i\alpha e^{i\varphi_2} \\
0 & 0 & c & 0 & i\alpha e^{i\varphi_2} & 0\\
0 & -i\alpha^\dagger e^{-i\varphi_1} & 0 & c & 0 & 0\\
-i\alpha^\dagger e^{-i\varphi_1} & 0 & -i\alpha^\dagger e^{-i\varphi_2} & 0 & c & 0\\
0 & -i\alpha^\dagger e^{-i\varphi_2} & 0 & 0 & 0 & c
\end{bmatrix},$
}
\end{equation}
\normalsize
where the frequency dependency enters through the $c=-i\omega +\kappa /2$ coefficient.
To express the intracavity modes through the vacuum input, we use the inverse matrix $\mathbf{M}^{-1}$:
\begin{equation}
\label{M-1}
    \tilde{a}(\omega)=\sqrt{\kappa}\mathbf{M}^{-1}\tilde{b}_{\text{in}}(\omega). 
\end{equation}
For the tripartite case, such a matrix is written in the following way 
    \small
\begin{eqnarray}
    \mathbf{M}^{-1}= \frac{1}{c^2-2\alpha^2} \nonumber  \cdot\\
    \resizebox{.9\linewidth}{!}{%
   $ \begin{bmatrix}
c-\frac{\alpha^2}{c} & 0 & \pmb{\frac{\alpha^2 e^{i\Delta \varphi}}{c}} & 0 & -i\alpha e^{i\varphi_1}& 0\\
0 & c & 0 & -i\alpha e^{i\varphi_1} & 0 & -i\alpha e^{i\varphi_2}\\
\pmb{\frac{\alpha^2 e^{i\Delta \varphi}}{c}} & 0 & c-\frac{\alpha^2}{c} & 0 & -i\alpha e^{i\varphi_2} & 0\\
0 & i\alpha e^{-i\varphi_1} & 0 & c-\frac{\alpha^2}{c} & 0 & \pmb{\frac{\alpha^2 e^{i\Delta \varphi}}{c}}\\
i\alpha e^{-i\varphi_1} & 0 & i\alpha e^{-i\varphi_2} & 0 & c & 0\\
0 & i\alpha e^{-i\varphi_2} & 0 & \pmb{\frac{\alpha^2 e^{i\Delta \varphi}}{c}} & 0 & c-\frac{\alpha^2}{c}
 \end{bmatrix},$
 }
\end{eqnarray}
\normalsize
where $\Delta\varphi=\varphi_1-\varphi_2$. Besides two mode squeezing (TMS) correlations proportional to $\alpha$, the matrix contains also  beamsplitter correlations (BS) $\propto \alpha^2$  (denoted in bold). Note that the $\alpha^2$-terms are absent in the matrix $\mathbf{M}$ introduced in Eq. (\ref{M}).
In this tripartite example, the phase difference $\Delta \varphi$ contributes only to the BS connection $\tilde{a}_i(\omega) \leftrightarrow \tilde{a}_j(\omega)$ or $\tilde{a}^\dagger_i(\omega) \leftrightarrow \tilde{a}^\dagger_j(\omega)$.
TMS connections are defined by entries $\tilde{a}_i(\omega) \leftrightarrow \tilde{a}^\dagger_j(\omega)$ in $\mathbf{M}^{-1}$.
In Appendix \ref{analytics} we discuss how the  phase shifts between pumps influence the structure of subspaces within the covariance matrix. 
In the quadripartite case, it turns out that those products of $\mathbf{M}^{-1}$ responsible for beamsplitter interaction can be suppressed fully by choosing pump phases properly in certain pump tone configurations.
\begin{figure}[ht]
\includegraphics[width=1\linewidth]{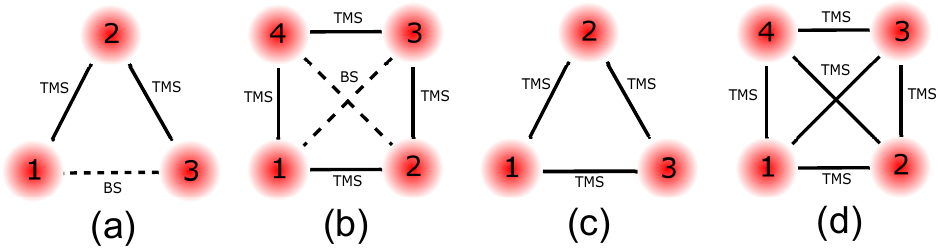}
	\caption{
	{\bf Graph representation for the entangled tripartite and quadripartite systems.}
	Generalized H-graph ($\mathscr{\tilde{H}}$-graph) representation of bisqueezed tripartite CV entangled state (a) and quadripartite CV entangled $\mathscr{\tilde{H}}$-graph state (b) obtained in our experiments. Vacuum modes (red circles) are connected via two-mode squeezing (TMS, solid lines on graph) and beamsplitter (BS, dashed lines) correlation. Graphs (c) and (d) represent tri- and quadripartite GHZ states, which can be obtained via introducing an additional pump tone with $\Delta_d=0$ in the tripartite case and two additional tones at $-\Delta$ and $\Delta$ in the quadripartite setting. The additional pumps supply missing TMS connections to the entangled states.  For details, see Appendix \ref{analytics}.}
	\label{Nmodegraph}
\end{figure}

Interestingly, the matrix $\sqrt{\kappa}\mathbf{M}^{-1}$ can also be interpreted as an \textit{adjacency} matrix, which is used in graph theory for characterization of the connections, the graph edges. In regular H(\textit{amiltonian})-graph \cite{Gu2009,menicucci2011graphical,bruschi2016towards,pfister2019continuous}, each vertex represents a vacuum mode and the adjacency matrix describes correlations produced by two mode squeezing (TMS) between the vacuum modes. 
However, our scheme produces additional correlations, BS correlations, that can no longer be described purely by the TMS correlations and, therefore, the standard H-graph theory needs a more generalized approach. 

In our approach, we introduce generalized $\mathscr{\tilde{H}}$-graphs formed by both two mode squeezing and beamsplitter correlations (see Appendix \ref{analytics}).
Examples of such graphs are presented in Fig.\,\ref{Nmodegraph}a, b for the tripartite and quadripartite case, respectively.
Intriguingly, the famous Greenberger-Horne-Zeilinger (GHZ) states \cite{briegel2001persistent,menicucci2011graphical,pfister2019continuous} have different structure since they are devoid of BS correlations.
However, by applying additional pump tones and adjusting phase difference between them, one can generate tripartite and quadripartite GHZ states consisting of only SQ correlations as shown in Fig.\,\ref{Nmodegraph}c,d.
In general, the considered multipumping scheme allows us to control SQ and BS bonds providing access to more complex structures of CV quantum states beyond GHZ-like states.

From the experimental point of view, the observer is interested in the adjacency matrix for out-coming modes $\tilde{b}_\text{out}$, which are obtained from $\tilde{a}$ using the input-output relationship in Eq. (\ref{IOeq}).
This equation yields
\begin{equation} \label{bout_th}
    \tilde{b}_{\text{out}}(\omega)=(\mathbf{I}-\kappa\mathbf{M}^{-1})\tilde{b}_{\text{in}}(\omega),
\end{equation}
on the basis of which we may define the adjacency matrix $\mathbf{\tilde{M}}= \mathbf{I}-\kappa\mathbf{M}^{-1}$ for input-output mode graphs. Due to the linear nature of the equation, the unit matrix does not change the intrinsic form of interactions between the vacuum modes, but the correlation structure of intracavity and output spectral modes is equivalent. Consequently,  analyzing a graph defined by the matrix $\mathbf{M}^{-1}$ is sufficient to characterize the BS and TMS connections between vacuum spectral modes.

\subsection{Connection to Hamiltonian graph}

CV cluster states with square-lattice graph structure provide a foundation for measurement-based continuous variable quantum computation (CVQC) \cite{menicucci2011graphical}. Cluster states can be asymptotically reached from H-graph states in the case of infinite squeezing \cite{bruschi2016towards}. 
The H-graph structure is defined by its adjacency matrix $\mathbf{G}$, whose entries $G_{ij}$ specify the multimode squeezing Hamiltonian as:
\begin{equation}
\label{Hamiltonian_multimodesqueezing}
    H_\text{S}= \hbar \alpha \sum_{i,j} G_{ij}(\tilde{a}_i^\dagger \tilde{a}_j^\dagger + \tilde{a}_i \tilde{a}_j).
\end{equation}
Here, the pump tone amplitudes are considered to have equal strength $\alpha$.
The matrix $\mathbf{G}$ involves TMS correlations between modes $\tilde{a}_i\leftrightarrow \tilde{a}^\dagger_j$, but as was pointed out before, the BS correlations do not show up in the H-graph representation.
The equations of motion for the operators are given by
$i\dot {\tilde a}_k^\dag  =\alpha \sum\nolimits_j {{G_{jk}}} {\tilde a_j}$ and 
$i{\dot {\tilde a}_k} =  -\alpha \sum\nolimits_j {G_{jk}^*} \tilde a_j^\dag $. Taking the Fourier transform, the left hand side equals $ \omega \times \tilde a(\omega)$, and the combination $\omega \tilde{a}_k(\omega)=\alpha \sum\nolimits_j {{G_{jk}}} {\tilde a_j}^\dag(-\omega)$ provides the connection to the QLE treatment in Eq. (\ref{M}): $\tilde a(\omega)$ here is the cavity signal 
defined by the graph connections given by $G_{jk}$. Consequently, the basic graph structures are the same, but the form of $\mathbf{M}^{-1}$ in the QLE analysis yields higher order correlations which are experimentally relevant.

A standard description for graphs is based on the complex symmetric matrix $\mathbf{Z}=ie^{-\mathbf{G}}$, which is interpreted as the adjacency matrix for an undirected  \textit{Gaussian} graph with complex-valued edge weights \cite{menicucci2011graphical,bruschi2016towards}. 
Decomposing such a matrix up to quadratic terms $\mathbf{Z}=i\mathbf{I}-i\mathbf{G}+\frac{i}{2}\mathbf{G}^2+\mathcal{O}(h^3)$, we obtain corrections to the adjacency matrix, which correspond to additional correlations, the BS correlations, obtained in our QLE analysis. 
Indeed, BS transformations embody interactions to second-order, which provides classical correlations between corresponding nodes \cite{bruschi2017entanglement}.

Let us now show the origin of BS correlations using the multimode squeezing Hamiltonian of Eq. \ref{Hamiltonian_multimodesqueezing} in $R = e^{-\frac{i}{\hbar} H \tau}$ where we have considered that the system is pumped for a finite time $\tau$.
The multimode squeezing operator $R$ can be decomposed to a combination of TMS operators, containing $B_{ij}=\tilde{a}_i^\dagger \tilde{a}_j^\dagger+\tilde{a}_i\tilde{a}_j$, and BS transformations based on $T_{ij}=\tilde{a}_i \tilde{a}_j^\dagger-\tilde{a}_i^\dagger\tilde{a}_j$.
By utilizing the Zassenhaus expansion (up to first order of commutation relationship) \cite{bruschi2017entanglement},  
we obtain for the tripartite squeezing operator
\begin{align}
    R=e^{-i\alpha\tau \sum^3_{i,j=1} G_{ij}(\tilde{a}_i^\dagger \tilde{a}_j^\dagger + \tilde{a}_i \tilde{a}_j)}=e^{-i\alpha \tau( B_{12}+ B_{23})}=e^{-i\alpha \tau B_{12}}\nonumber\\
     e^{-i\alpha  \tau B_{23}}e^{\frac{\alpha^2 \tau^2 }{2}[B_{12},B_{23}]}=
    e^{-i\alpha  \tau  B_{12}}e^{-i\alpha  \tau  B_{23}}e^{\theta_{13}T_{13}}.
    \label{ZahenhaussExpansion}
\end{align}
Here, $\theta_{13}$ specifies the relative phase shift between the two pump tones. For detailed information on the expansion coefficients for a bisqueezed state we refer the reader to Ref. \onlinecite{bruschi2017entanglement}.

The total multimode squeezing operation can be considered as a combination of TMS operators, acting on the respective bipartitions, and BS transformations between the other modes. 
The beamsplitter correlations are phase dependent, and the strength of the BS contribution can be tuned down to zero in certain cases via proper choice of the phase difference between the pumps.

The general decomposition of the multimode squeezing operator can be expressed as
\begin{equation}
    R= \prod_{1}^{N_\text{TMS}}e^{-i\alpha \tau B_{12}}e^{-i\alpha \tau B_{23}}\dots\prod_{1}^{N_\text{BS}}e^{\theta_{13} T_{13}}e^{\theta_{24}T_{24}}\dots,
\end{equation}
where the total number of TMS and BS operators in the decomposition is
   $ N_{\text{BS}}=\sum_{n} N-2n$ and $
    N_{\text{TMS}}=\sum_{n} N-2n+1$;
where $n=1,2,\dots,\lfloor\frac{N}{2}\rfloor$
for the configuration introduced in Fig.\,\ref{modes_separation_fig} with $N-1$  pump tones and $N$ spectral modes.
Collecting all of the entries of $B_{ij}$ and $T_{ij}$, we obtain a \textit{generalized} adjacency matrix $\mathbf{\tilde{G}}$ with entries $\tilde{G}_{ij}=\sum_1^{N_{\text{TMS}}}B_{ij}+\sum_1^{N_{\text{BS}}}T_{ij}$.  Thus, the beamsplitter correlations in adjacency matrix arise naturally from the squeezing operator formalism when the Hamiltonian is supplied with the second order terms in pump amplitude. The structure of the matrix $\mathbf{\tilde{G}}$ defines the edge connections in the \textit{generalized} $\mathscr{\tilde{H}}$-graph.

\subsection{Verification of the multipartite entanglement}\label{verification}

The generalized graph analysis allows us to visualise the structure of entanglement in the  quantum state generated by simultaneous multiple pump tones.
However, in order to estimate the amount of quantum resources embedded in the state, we have to investigate and quantify the classical and quantum correlations and determine how they reflect the genuine  multipartite entanglement of the state.

Within the framework of parametric amplifiers, all microwave fields produced by a JPA below the critical threshold are Gaussian \cite{braunstein2005quantum, Adesso2014}.
Therefore, the output states of a $N$-mode JPA  can be fully characterized by its covariance matrix of $2N$-length column vector with quadratures $\tilde{r}~=~\left(\tilde{x}_1, \tilde{p}_1, \dots \tilde{x}_N, \tilde{p}_N \right)^{T}$, where $ \tilde{x}_i = (\tilde{a}_i+\tilde{a}^{\dagger}_i)/2$ and $\tilde{p}_i=(\tilde{a}_i-\tilde{a}^{\dagger}_i)/2i$.
The covariance matrix $\mathscr{V}$, whose elements are given by
\begin{equation} \label{scaledCov}
    \mathscr{V}_{ij}=2\left<\Delta\tilde{r}_i \Delta\tilde{r}_j + \Delta\tilde{r}_j \Delta\tilde{r}_i \right> - 4\left<\Delta\tilde{r}_i \right> \left<\Delta\tilde{r}_j\right>,
\end{equation}
is sufficient for detection of the entanglement, eliminating the need for analysis of the full density matrix.
The last term can be ignored as we take $\Delta \tilde{r}_i=\tilde{r}_i-\left<\tilde{r}_i\right>$.
Obtaining the covariance matrix, we can analyse the entanglement and examine the structure of the quantum state.  
In this work, we consider fully inseparable states and genuinely entangled states, for which the covariance-based detection is, in general,  more robust than detection via complete determination of the state  \cite{Loock2003Detecting}. 
While the covariance matrix is sufficient for evaluating entanglement of Gaussian states, it is necessary to include higher-order correlations in the evaluation of non-Gaussian states \cite{Zheng2010, Sabin2020}.

To examine inseparability properties of the quantum state \cite{Simon2000,Duan2000,Mancini2002,Loock2003Detecting,Hyllus2006,Teh2014criteria,Shchukin2015}, we apply symplectic transformations to the covariance matrix and calculate its symplectic eigenvalues -- the PPT criteria \cite{Peres1996, horodecki1996}. 
Such transformations are equivalent to a phase space reflection of a single party in the $N$-partite state \cite{Simon2000}.
All minimum symplectic eigenvalues $\{\nu_i\}_{i=1}^N$ would be less than one, which indicates that this partially time-reversed state is unphysical; in other words, the original state is fully inseparable.
As has been pointed out in Ref. \onlinecite{Teh2014criteria}, if the purity of states cannot be guaranteed in an experimental setting, verification of full inseparability in a multimode system does not imply genuine multipartite entanglement (GME).

The entanglement structure becomes more involved with increasing number of parties.
While the symplectic transform approach indicates that any one partite were inseparable from the whole, a state that is a mixture of separable states would show full inseparability based on this PPT criterion.
The states that cannot be written in such a way are called {\it genuinely} entangled \cite{braunstein2005quantum} and the verification of such states differs from full inseparability.
Using generalized position and momentum observables, an entanglement criterion has been derived  and applied to confirm tripartite energy-time entanglement of three spatially separated photons \cite{shalm2013three}. 
In particular, there is an universal GME criterion derived in Ref.\,\cite{Loock2003Detecting} and further refined in Ref.\,\cite{Teh2014criteria}. This GME criterion utilizes only variances of  quadrature operators and it can be used for entanglement verification without any additional measurements.
This general criterion was recently employed for verification of genuine tripartite entanglement of  microwaves in a double superconducting cavity setting  \cite{Wilson2018generatingmultimode}.

The GME criterion is based on the weighted variance of the quadratures, $u = \sum_i h_i \tilde{x}_i$ and $v = \sum_k g_k \tilde{p}_k$; $i,k=\left\lbrace 1,2,3,\dots, N \right\rbrace$.
Violation of the inequality
\begin{eqnarray}
	\label{Scrit_3}
	S \equiv \frac{\left< \Delta u^2 \right> + \left< \Delta v^2 \right>}{f_3(h_i,g_i)} \geq 1,
\end{eqnarray}
where
\begin{eqnarray}
	f_3(h_i,g_i) =  \frac{1}{2} \min \lbrace & |h_1 g_1 + h_2 g_2| + |h_3 g_3|, \nonumber \\
	& |h_3 g_3 + h_2 g_2| + |h_1 g_1|, \nonumber \\
	& |h_1 g_1 + h_3 g_3| + |h_2 g_2|, \rbrace  \nonumber
\end{eqnarray}
is sufficient to confirm genuine tripartite entanglement ($N=3$) and violation of

\begin{eqnarray}
	\label{Scrit_4}
	S \equiv \frac{\left< \Delta u^2 \right> + \left< \Delta v^2 \right>}{f_4(h_i,g_i)} \geq 1,
\end{eqnarray}
where
\begin{eqnarray}
	f_4(h_i,g_i) =  \min \lbrace & |h_1 g_1 + h_2 g_2 + h_3 g_3| + |h_4 g_4|, \nonumber \\
	& |h_4 g_4 + h_2 g_2 + h_3 g_3| + |h_1 g_1|, \nonumber \\
	& |h_4 g_4 + h_1 g_1 + h_3 g_3| + |h_2 g_2|, \nonumber \\
	& |h_4 g_4 + h_1 g_1 + h_2 g_2| + |h_3 g_3|, \nonumber \\
	& |h_1 g_1 + h_2 g_2| + |h_3 g_3 + h_4 g_4|, \nonumber \\
	& |h_1 g_1 + h_3 g_3| + |h_2 g_2 + h_4 g_4|, \nonumber \\
	& |h_2 g_2 + h_3 g_3| + |h_1 g_1 + h_4 g_4|\rbrace  \nonumber
\end{eqnarray}
is sufficient to confirm genuine quadripartite entanglement ($N=4$) with weights $h_i,g_k$ being in range $[-1,1]$. 
To simplify the search domain, we set $h_1=g_1=1$ and $h_i = h, g_i = g$, for $i=\left\lbrace 2,3 \right\rbrace$ or $i=\left\lbrace 2,3,4 \right\rbrace$ with respect to the number of parties $N$ \cite{Teh2014criteria}.

In the double and triple pumping schemes, we generate generalized tripartite and quadripartite $\mathscr{H}$-graph states, which have a different structure compared with Greenberger-Horne-Zeilinger (GHZ) states \cite{briegel2001persistent,menicucci2011graphical,pfister2019continuous}.
The generalization deals with addition of BS correlations in the $\mathscr{H}$-graph structure.
However, by applying additional pump tones and adjusting the phase difference between them, we can obtain regular GHZ type of entangled states.
Thus, our scheme facilitates control of TMS and BS correlations and, thereby, allows tuning of the structure of the entangled state.

Typically, the experimental weights of the graph edge connections are slightly non-symmetric due to imperfections in the measurement settings. This results in a
difference in the optimal weight values in the GME criterion. 
In order to find the full violation of the criterion in our analysis, we swap over all possible "base" modes (with weights $h_i=g_i=1$) in order to detect the minimum value for $S$.

\section{Experiment}
\label{Experiment}
\subsection{Experimental methods}
\begin{figure}
\includegraphics[width=1\linewidth]{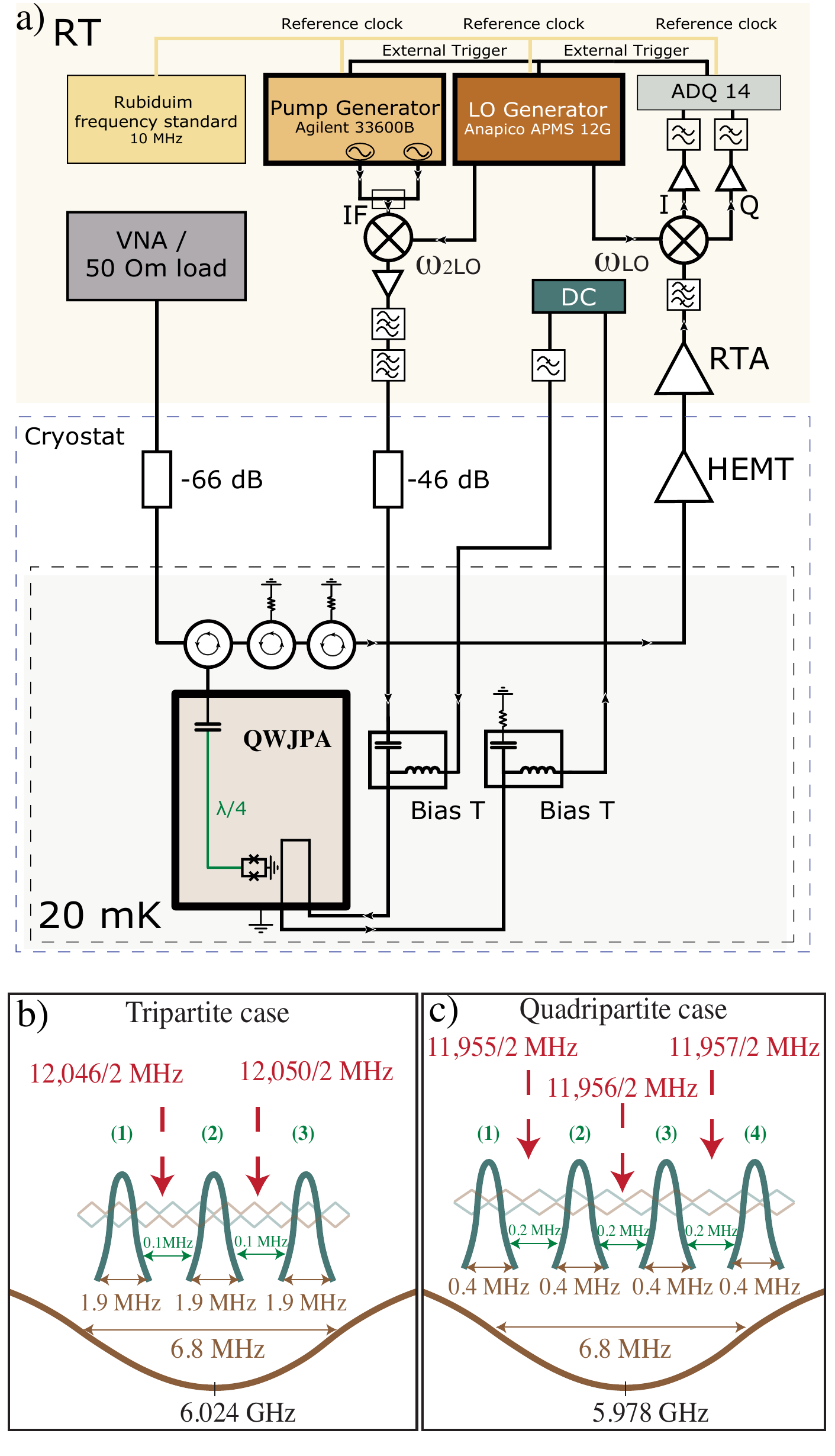}
	\caption{
	{\bf Experimental scheme and device characterization.} 
	(a) Principle of the experimental setup for tripartite entanglement measurements. The device is connected to the test ports via circulators. The DC bias current and AC pumping of flux are combined and reseparated  in bias-T components. Depending on the measurement type, input is connected either to a vector network analyzer (VNA) or a 50 $\Omega$ termination, whereas the output is directed either to a VNA, a signal analyzer, or a analog/digital converter. The frequency span of spectral modes and their separation is given for tripartite and  quadripartite case in frames (b) and (c), respectively.  }
	\label{expFig}
\end{figure}

In our microwave experiments, we employ a niobium $\lambda/4$ coplanar $50\,\Omega$ transmission line terminated into a SQUID loop (QWJPA), forming a quarter-wave Josephson parametric amplifier.
The SQUID's junctions are formed by Nb/Al-Al$_2$O$_3$/Nb 1$\times$1 $\mu$m tunnel barriers with $I_c\simeq 4$ $\mu$A critical current. 
The JPA, operating in three-wave-mixing mode around the cavity frequency $\omega_r$, is pumped by an external RF magnetic flux through the SQUID loop using a single turn pump coil at frequency $2\omega_r$ (marked as $2\omega_{\mathrm{LO}}$ in Fig.\,\ref{expFig}a) \cite{yamamoto2016parametric, Elo2019}. 
We chose this operation regime, because for four-wave mixing (typically using a current pump near $\omega_r$), the large amplitude pump is within the amplification bandwidth, whereas the three-wave mixing process separates the pump tone from the amplified signals, thus simplifying the practical use of the JPA.
The loaded quality factor at the operation frequency is $\sim$900, while the internal $Q$ is by a factor of three larger.
The basic (zero-flux) resonator frequency is 6.115\,GHz, it can be tuned below 5.5\,GHz by imposing external DC magnetic flux through the SQUID.

Our measurement setup is illustrated in Fig.\,\ref{expFig}a. 
The experiments were conducted at 20 mK using a BlueFors LD400 dry dilution cryostat. 
The JPA was protected from external magnetic fields using a Cryoperm shield. 
The DC flux bias and the RF pump shared a common on-chip flux line, and the signals were combined in an external bias-tee.
Since our basic microwave setting is for reflection measurements, the sample is connected to the input and output ports via a circulator having a frequency band of $4-12$\,GHz.
A vector network analyzer (VNA) was used to characterize the sample, whereas during the entanglement generation measurements, the signal port was kept terminated.
By applying a multitone pump to JPA, correlated microwaves are generated from vacuum fluctuations. In the tripartite setting illustrated in Fig.\,\ref{expFig}a, we had control over the relative phases of the pumps directly whereas, in the quadripartite case, phase rotation was possible in the data analysis only.
Basic experimental data in the tripartite case as well as determination of the cavity parameters $\kappa$, $\gamma$ and $K$ are discussed in Appendix \ref{systpara}.

\paragraph{Tripartite case.} 
In the tripartite case, phase-controlled pump signals from the RF waveform generator are mixed with the frequency-doubled local oscillator (LO) frequency, filtered by a pair of home-made, tunable bandpass cavity filters. In order to avoid spurious pumping at $\omega_{\mathrm{2LO}}$, a band rejection filter tuned to $2\omega_{\mathrm{LO}}$ is employed.
The filtering ensures passage only for the desired two pump signal at angular frequencies $\omega_1=\omega_r-\Delta/2$ and $\omega_2=\omega_r+\Delta/2$.
Sufficient noise thermalization is ensured by the 46\,dB attenuation because the pump coil is only weakly coupled to the SQUID. 

We apply a DC magnetic flux of $\Phi_{DC}=0.383\Phi_0$ through the SQUID loop, resulting in $\frac{\omega_r}{2\pi}=6.024$\,GHz for the cavity frequency.
In the three-mode experiment, we apply two pump tones at  $(2\times\frac{\omega_r}{2\pi}-2)$\,MHz and $(2\times\frac{\omega_r}{2\pi}+2)$\,MHz, with the half-frequencies positioned as depicted in Fig.\,\ref{expFig}b and the correlated spectral modes are defined symmetrically with respect to the pump half-frequencies. 
Each mode has a bandwidth of $1.9$\,MHz and is separated from the other modes by $0.1$\,MHz.
The phase control in the measurement is facilitated by phase-locking of microwave generators to a 10\,MHz Rubidium reference clock and by using a joint external trigger.

To collect data for correlation analysis, we mix down the output signal using a synchronized LO signal and record the output quadratures using  two channels of a Teledyne SP Devices ADQ14 digitizer with sample rate of 50\,MSa/s per channel covering the bandwidth of 25\,MHz.
Furthermore, we employ an overall detuning of 14\,MHz, i.e. a heterodyne detection scheme, in order to avoid $1/f$ noise from the measurement devices and the IQ mixer in the frequency conversion part of the setup.
Using digital postprocessing, we can easily shift the center frequency of the heterodyned MHz signal to zero, ready for final correlation analysis of the modes. 

 Our three-mode measurement scheme provides the remarkable advantage of physical control of the phase difference between the two pump tones, which is essential for the analysis of phase dependence in the entangled states.
The phase difference between 2\,LO signal used as the carrier of the pump tones and LO readout signal remains fixed.
Therefore, a change of the initial phase of the IF-signal in one of the pump generator channels (Agilent 33600B in Fig.\,\ref{expFig}a) relative to the second IF channel creates an effective phase difference $\Delta\varphi$ between two pump tones.
Importantly, this difference $\Delta\varphi$ is preserved after mixing with 2LO (using simplified notation): $e^{-i(\omega_1 t + \Delta\varphi)}e^{-i\omega_{\text{2LO}} t}=e^{-i((\omega_1 +\omega_{\text{2LO}})t+\Delta\varphi)}; e^{-i(\omega_2 t)}e^{-i\omega_{\text{2LO}} t}=e^{-i((\omega_2 +\omega_{\text{2LO}})t)}$.

Since our fully phase-locked scheme preserves  the phases of the received, demodulated output quadratures, the measurement of covariance matrix components can be averaged for reducing noise in the elements.
Finally, the reference phase of a single mode (defining the basis for I and Q) can be adjusted in postprocessing step in such a way that the corresponding subspace of the covariance matrix becomes a diagonal $2\times2$ matrix.

In the tripartite case, indeed, we find that the hardware-controlled relative phase rotation (in addition to the reference phase value to both channels of the pump generator) is equivalent to a proper phase rotation in the postprocessing step.
The postprocessing will be discussed in more detail in Section\,\ref{entanglement_results}.\\

\paragraph{Quadripartite case.}
In the four spectral mode case, we simplify the experimental setup by eliminating the physical phase control, and replaced it by postprocessing of the received signals.
This simplification possibility highlights the scalability of our entanglement generation method.
The employed digital postprocessing is equivalent to hardware-level selective separation of spectral modes into four channels, e.g. using bandpass filters in conjunction with power splitters, and additional tunable delay lines for each selected spectral mode frequency. 

We apply a DC magnetic flux of $\Phi_{DC}=0.417\,\Phi_0$ through a SQUID resulting in $\omega_r=5.978$\,GHz cavity frequency that slightly differs from the tripartite case, see Fig.\,\ref{expFig}b,c.
In order to generate quadripartite correlations, we apply three pump tones using Anapico APMS 12G generator, using strong high-pass filtering (2 of Mini-Circuits VHF-8400+) to avoid subharmonic transmission to the circuitry. 
In this scheme, we avoid any external mixers for the input pump microwaves.
By applying three phase-locked pump tones at frequencies $2\times\frac{\omega_r}{2\pi}$\,MHz, $(2\times\frac{\omega_r}{2\pi}+1)$\,MHz, and $(2\times\frac{\omega_r}{2\pi}-1)$\,MHz, we generate  four correlated spectral modes out from the ground state of the microwave cavity.
Each mode has a bandwidth of $0.4$\,MHz and is separated from adjacent modes by $0.2$\,MHz.
The output microwaves are captured, mixed down and digitized by Anritsu MS2830A Signal Analyzer with a bandwidth of $2$\,MHz. Again, averaging is needed to lower the noise in the covariance matrix elements, and in this scheme, digital postprocessing is necessary to unify the phase settings in the covariance matrices before summation.

The experimental detection of $\mathcal{\tilde{H}}$-graph states and their genuine multipartite entanglement depends on relations among the covariance matrix elements as discussed in Section \ref{verification}. The degree of violation in the GME condition $S<1$ depends strongly on the magnitude ratio of the diagonal covariance elements to the off-diagonal ones. Therefore, calibration of the detected signal powers is decisive, which is discussed in Appendix \ref{systgain}.

\subsection{Scaled covariance matrix}
The system gain $\left<G_{\Sigma}\right>$ determined in Appendix \ref{systgain} refers to measured power per unit band width.
Since the measured spectral mode quadratures $I_i$ and $Q_i$ are determined over the band $\Delta f_i$, the scaled quadrature $x_i$, equivalent to the amplitude of the quantum mechanical operators  $\mathbf{x}$, is given by the formula
\begin{equation}
	{x}_i=\frac{I_i}{\sqrt{G_i Z_0 h f_i \Delta f_i}},
\end{equation}
where $G_i=\left<G_{\Sigma,i}\right>$ is the system gain for  $i^{\mathrm{th}}$ spectral mode, $Z_0 = 50$\,$\Omega$ is transmission line impedance and $\Delta f_i$ is the bandwidth of the spectral mode: $\Delta f_i = 2$ MHz or $\Delta f_i = 0.4$ MHz for tripartite and quadripartite case, respectively. Similar scaling is applied also to the quadrature component $Q_i$.

Similar to our earlier work \cite{lahteenmaki2013dynamical}, the  noise added by the preamplifier is subtracted from the diagonal elements of the covariance matrix (see Eq. \ref{scaledCov}):
\begin{equation}
	\mathscr{V}=4\left( \mathbf{V}_{\text{on}}-\mathbf{V}_{\text{off}} \right) + \mathbf{I} \coth\frac{hf_i}{2 k_b T_i}.
\end{equation}
where $\mathbf{V}_{\text{off}}$ denotes the covariance matrix measured in the absence of the pump. Due to scaling of the covariance matrix $\mathscr{V}$, this equation yields a unity diagonal matrix in the absence of pumping at $T \rightarrow 0$.
The average physical temperature in our experiments is $T_i=20$\,mK resulting in $\coth\frac{hf_i}{2 k_b T_i}=1.000$.

\section{Multipartite entanglement}
\label{entanglement_results}

To characterize the structure of the entanglement in output states, we analyze the resulting covariance matrices using positive partial transpose (PPT) formalism and GME criteria discussed in Section \ref{verification} for tripartite and quadripartite cases. \\

\paragraph{Tripartite case.} \label{tripartite}

Leveraging the amplitude and phase control of the pump signals, we experimentally evaluate the PPT and GME criteria values at different pump parameters.
For comparison, we also conducted detailed numerical simulations based on the QLE in Eq. \ref{QLE} using experimentally determined JPA parameters in the measurements. In general, we find good agreement between simulations and the experimental data, which is reassuring concerning the validity of the results. 
\begin{figure*}[tbh]
	\includegraphics[width=0.95\linewidth]{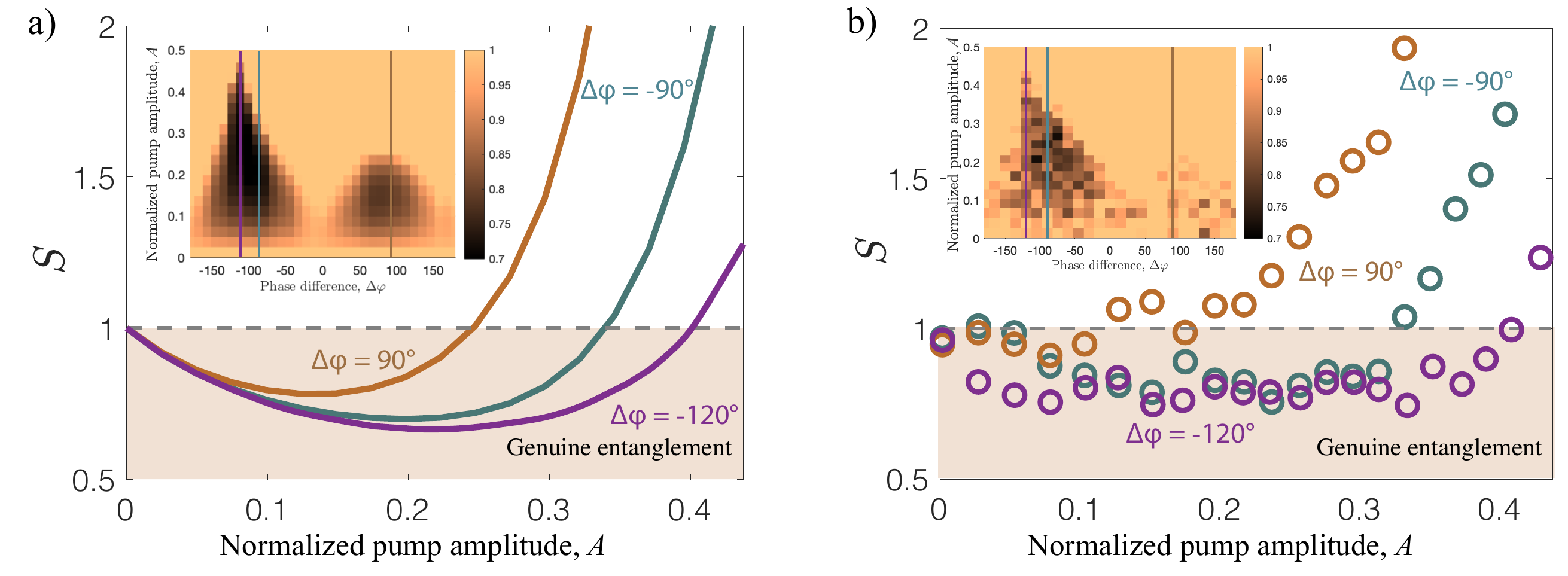}
	\caption{
	{\bf Phase-dependent genuine entanglement of tripartite bisqueezed state.} 
	(a) Simulation results for GME criterion as a function of normalized pump amplitude for three different values of the phase difference $\Delta\varphi$ between pump signals indicated in the figure; the simulation parameters were set to match the experiment (see Appendix \ref{systpara}). 
	The inset illustrates $S(A,\Delta\varphi)$ up to critical amplitude $A=0.5$. The weights $h_i, g_i$ were optimized in the calculation of $S$ as discussed in the text.
	(b) Experimental values for  
	$S$ 
	as a function of $A$ at the same phase difference values $\Delta\varphi$ between pump signals as in frame (a).	The inset illustrates measured $S(A,\Delta\varphi)$ up to  $A=0.5$.
	In general, the measured $S(A,\Delta\varphi)$ corresponds quite well to the inset in frame (a). Due to noise, the measured GME nearly vanishes around $\Delta\varphi \simeq +90^\circ$ where even the simulated $S$ is only slightly below 1. 
	The best genuine multipartite entanglement is reached at $\Delta\varphi = -90^\circ \dots -120^\circ$ owing to phase shifts introduced by the cavity (see Fig. \ref{cav_phase} in Appendix \ref{phase_resp}). The parametric drive changes the phase response of the cavity which leads to a shift in the optimum conditions for GME as a function of $A$.
	}
    \label{3modes}
\end{figure*}

\begin{figure}[htb]
\includegraphics[width=1\linewidth]{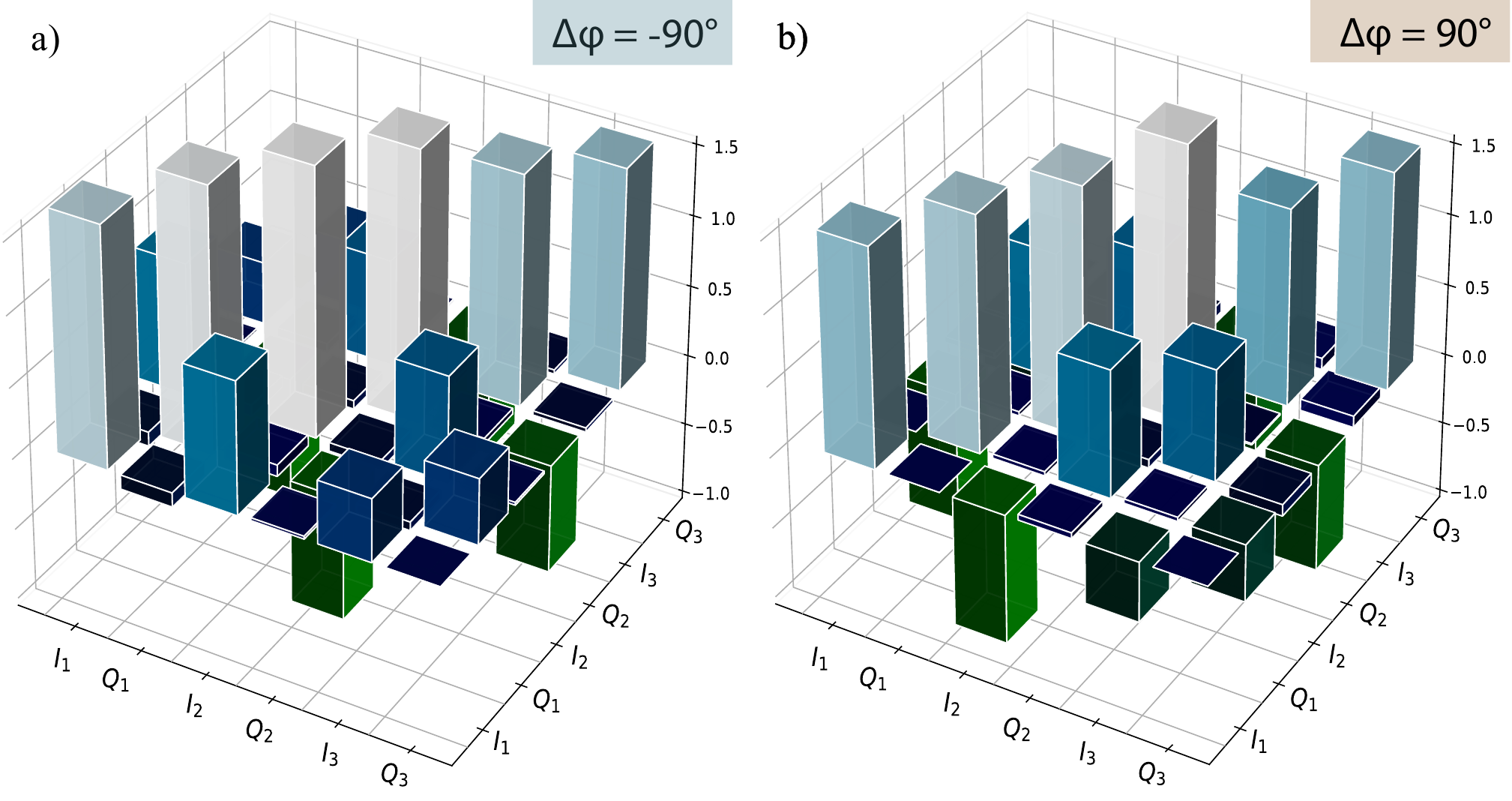}
	\caption{
	{\bf Covariance matrix of the genuinely entangled tripartite bisqueezed state.}
	Experimentally obtained tripartite covariance matrices for  genuinely entangled bisqueezed states.  The phase difference between the two displayed cases is $180^\circ$. Via this control of pump phases we demonstrate the rotation of the desired subspace elements, corresponding to TMS type of correlations ($1-2$ or $2-3$). Subspace elements related to BS correlations ($1-3$), shows always dependency on the distribution of the elements in the corresponding driven TMS subspaces. However, the eigenvalues of the BS subspace remain constant.} 
	\label{3_modes_result}
\end{figure}

Fig.\,\ref{3modes} depicts our experimental results on genuine tripartite entanglement and their comparison with simulations.
Fig.\,\ref{3modes}a illustrates results of numerical simulations on GME in terms of $S$ defined in Eq. \ref{Scrit_3}. At weak pumping, the condition for genuine entanglement $S < 1$ is fulfilled almost independent of the pump phases, but with increasing $A$, the simulations reveal an even smaller range of $\Delta \varphi$ yielding $S < 1$ (see the inset in Fig.\,\ref{3modes}a). The strongest genuine tripartite entanglement is reached at $\Delta \varphi \simeq -120^{\circ}$ under normalized pumping amplitude $A \simeq 0.22$, at which the simulations reach $S=0.70$. At the minimum of $S$, the corresponding weights are $h_i=\lbrace1,-0.65,-0.65\rbrace$ and $g_i=\lbrace1,0.65,0.65\rbrace$.
It is noteworthy that the phase setting $\Delta \varphi = + 90^{\circ}$ yields clearly worse entanglement than $\Delta \varphi = -90^{\circ}$. This asymmetry in GME between $\Delta \varphi = \pm 90^{\circ}$ arises from differences in the covariance matrices which is illustrated in Fig. \ref{3_modes_result}. 

Our experimental data on $S$ in Fig.\,\ref{3modes}b displays similar features as Fig.\,\ref{3modes}a. The measured GME criterion $S$ as function of normalized pump amplitude for three phase differences is shown in Fig.\,\ref{3modes}b. In the experimental data, nearly no GME is observed at positive phase differences, whereas $\Delta \varphi = - 90^{\circ}$ and $\Delta \varphi = - 120^{\circ}$ yield suppression down to $S=0.75 \pm 0.05$. The measured result at $\Delta \varphi = - 120^{\circ}$ follows quite well the simulated behavior as a function of the pump amplitude, and genuine entanglement is observed in the normalized amplitude range $A \in [\sim 0.01, 0.4]$. Overall, the pattern of $S(\varphi, A)$ in the inset of Fig.\,\ref{3modes}b coincides with the simulated pattern in Fig.\,\ref{3modes}a. The agreement strongly supports the presence of genuinely entangled bisqueezed state in our experiment. We emphasize that the optimum entanglement at $\Delta \varphi = - 120^{\circ}$, observed both in our simulations and in the experiment, cannot be obtained from  a simple analytical calculations for the lossless, strongly coupled model. The reason is the frequency-dependent phase response of the cavity due to finite coupling and dissipation rates (see Section \ref{phase_resp}), which, when included in the simulations, result in very good matching with the experiment.

Covariance matrices measured at the pump phase difference $\Delta \varphi = + 90^{\circ}$ and $\Delta \varphi = - 90^{\circ}$ are illustrated in Fig. \ref{3_modes_result}.
Technically, by rotating the phase of a pump signal, we selectively control certain subspace of the covariance matrix, which can be seen in Fig.\,\ref{3_modes_result}. An applied phase shift to the 1st or the 2nd pump rotates directly the subspace corresponding to two mode squeezing correlations, modes $1-2$ or $2-3$, respectively. If no phase shift to the selected pump tone is applied, the corresponding TMS subspace preserves its distribution of covariances. The subspace spanned by modes $1$ and $3$, corresponding to the beamsplitter type of correlations, has a structure according to products of the involved TMS subspaces. Distinct control of the BS subspace alone (leaving the TMS subspaces fixed) using a rotation of the pump phases is not possible.

Comparison of Figs. \ref{3_modes_result}a and \ref{3_modes_result}b, reveals how the subspaces transform with the phase difference from $\Delta \varphi = - 90^{\circ}$ to $\Delta \varphi = + 90^{\circ}$. Subspace $1-3$ corresponding to BS correlations shows a sign inversion in its elements. Subspace $2-3$ corresponds to the pump, the phase of which has not been changed and, thus, its elements remain fixed. The phase of the first pump has been changed by $\pi$ which inverts the TMS correlation in subspace $1-2$. 
These subspaces which are controlled by the pump phase settings, can  expressed in symmetric $\left<I_1I_2\right> \simeq \left<I_2I_3\right>$ or antisymmetric form $\left<I_1I_2\right> \simeq -\left<I_2I_3\right>$, see Figs. \ref{3_modes_result}a and \ref{3_modes_result}b, respectively.

\begin{figure}[!tb]
	\centering
	\includegraphics[ width=1\linewidth]{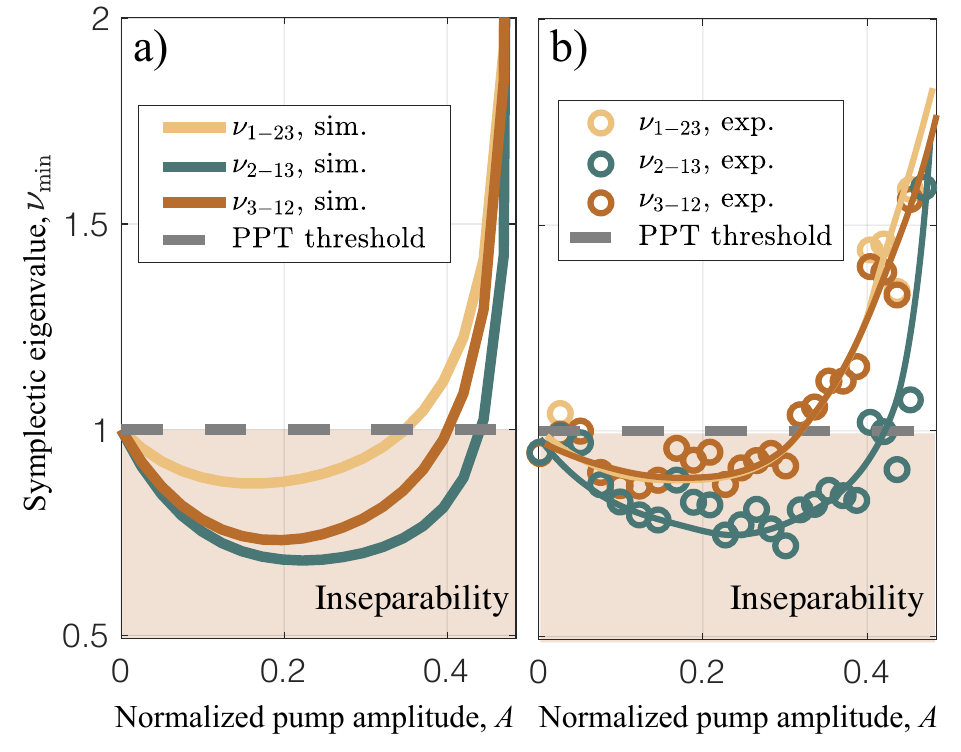}
	\caption{
	{\bf Phase-independent full inseparability of tripartite bisqueezed state.} 
a) PPT criteria in terms of the minimum symplectic eigenvalues simulated for our double-pump QWJPA using experimentally determined parameters. Eigenvalues $\min\lbrace\nu_i\rbrace$ are traces over normalized pump amplitude $A$; permutations ${1-23}$, ${2-13}$ and ${3-12}$ have been considered. The symplectic eigenvalues are the smallest for time-reversed second mode ($\nu_{2-13}$), which participates to both TMS processes.
b) Experimentally determined symplectic eigenvalues for the same permutations (\textcolor{yellow}{$\circ$},\,\textcolor{green}{$\circ$},\,\textcolor{red}{$\circ$}); the solid lines are just to guide the eyes. We find full inseparability at normalized pumping $0.05\lesssim A \lesssim 0.3$ in the experiment. Grey dashed line displays the full inseparability threshold. The difference in the simulated behavior of  $\nu_{1-23}$ and $\nu_{3-12}$ is caused by asymmetry due to finite value of resonance detuning $\Delta_r$.}
	\label{PPT3modeSlice}
\end{figure}

The inseparability of the covariance matrix was investigated using the PPT criteria (see Sect. \ref{theorFound}). Symplectic eigenvalues obtained for mode partitions $1-23$, $2-13$, and $3-12$ are depicted in Fig.\,\ref{PPT3modeSlice}a for simulations, while experimental data are displayed in Fig.\,\ref{PPT3modeSlice}b; here the first index specifies the mode in which the sign of the momentum has been reversed. The results are plotted as a function of normalized pump amplitude since the phase difference between the pump tones does not play a role.
Indeed, while the genuine entanglement is sensitive to both pump amplitude and phase difference, the PPT criterion is phase independent -- the minimum symplectic eigenvalues remain constant when the phase difference is varied at fixed pump amplitude.
Therefore, by exercising phase control over each pump, we gain the ability to switch from fully inseparable state to genuinely entangled state, without making any changes to the type of interaction between the modes. According to experimental results in  Fig.\,\ref{PPT3modeSlice}b, the middle frequency acted on by both pumps is the most inseparable part of the covariance matrix.

\begin{figure}
	
	\centering
	\includegraphics[ width=1\linewidth]{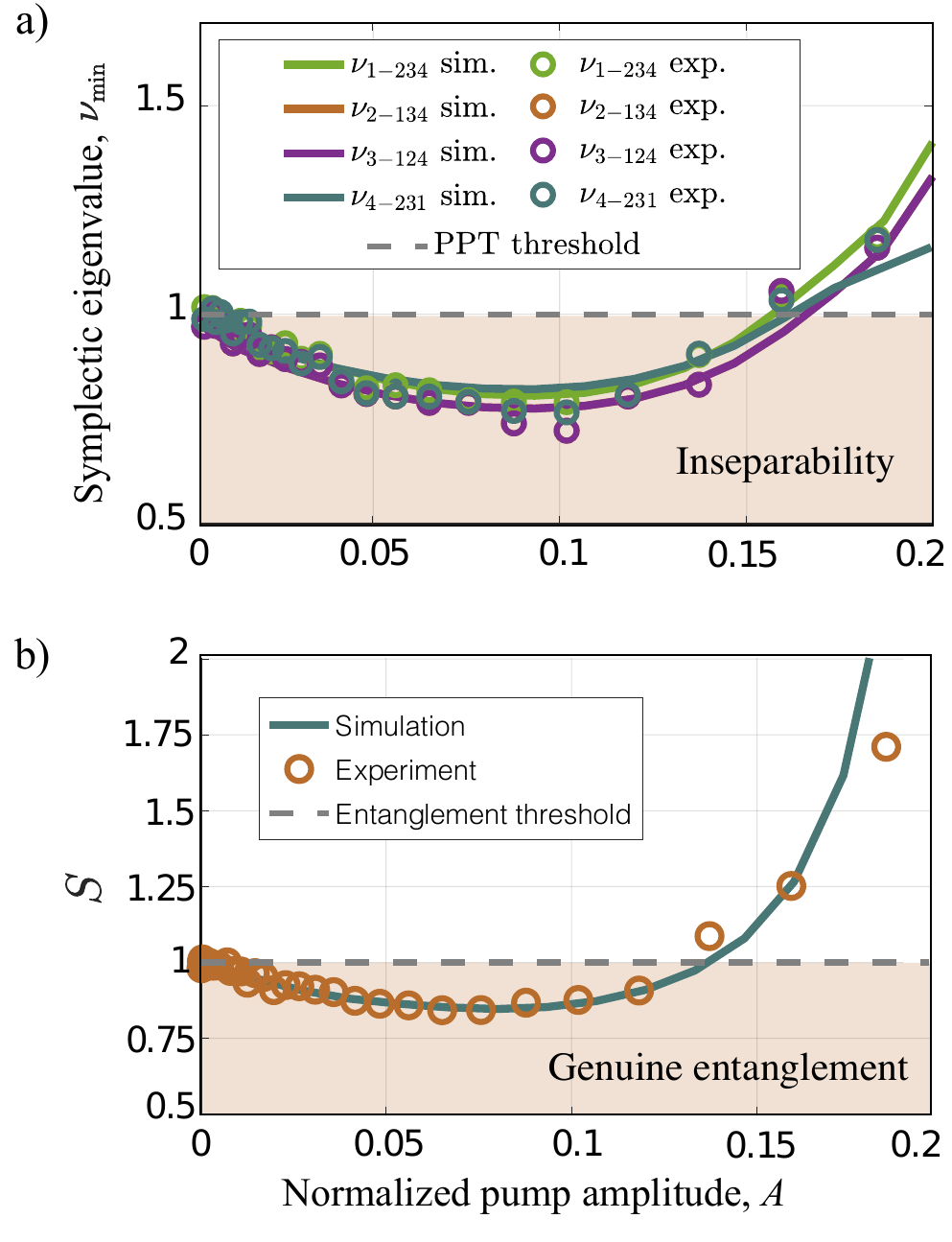}
	\caption{	{\bf Full inseparability and genuine entanglement of quadripartite $\mathscr{\tilde{H}}$-graph state (generalized H-graph)}. 
	a) Results on PPT criterion for four permutations of a four-mode Gaussian state.  Minimum eigenvalues $\min\lbrace\nu_i\rbrace$, indicated as open circles, are traced over normalized pump amplitude $A$. Results of our QLE simulations, plotted as solid curves, exhibit good correspondence with the experimentally obtained values. The full inseparability condition $\lbrace\nu_{1-234}<1\rbrace \bigwedge \lbrace\nu_{2-134}<1\rbrace \bigwedge \lbrace\nu_{3-124}<1\rbrace \bigwedge \lbrace\nu_{4-123}<1\rbrace$ is fulfilled in the range of $0.01\lesssim A\lesssim 0.15$. The green dashed line displays the entanglement threshold.}
	\label{PPT}
\end{figure}
\begin{figure}
\includegraphics[width=1\linewidth]{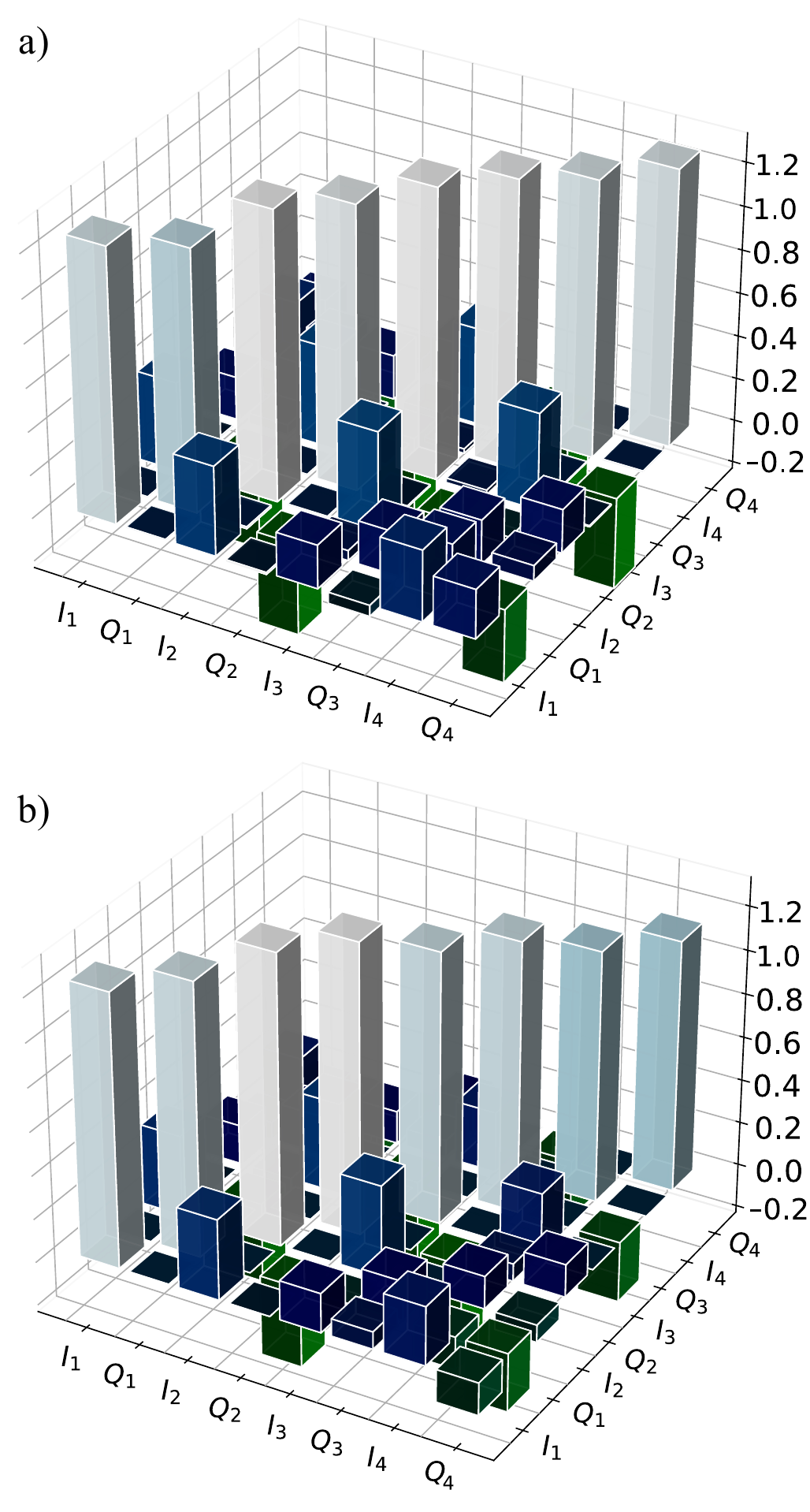}
	\caption{
	{\bf Covariance matrix of genuinely entangled quadripartite $\mathscr{\tilde{H}}$-graph} state. 
	a) Experimental covariance where the rotation of the TMS subspaces 1-2, 2-3, amd 3-4 have been made in such a way that the structure coincides with the matrix in Eq. (\ref{covarianceEq4modes}) of Appendix \ref{analytics} (each pump has phase $\frac{\pi}{2}$). The employed pump amplitude $A\eqsim 0.08 $ yields the smallest value for $S$.  b) Simulated covariance matrix using equal pump phases $\frac{\pi}{2}$ at $A\eqsim 0.08 $. The difference from the matrix in Eq. (\ref{covarianceEq4modes}) is due to the cavity response that induces extra phase shifts.
	}
	\label{CovMat_exp}
\end{figure}

\paragraph{Quadripartite case}
For the quadripartite case, we apply three pump drives with identical amplitude $\alpha_1=\alpha_2=\alpha_3=\alpha$.
While our goal is to demonstrate genuine entanglement generation of cluster states (mode structure depicted on Fig. \ref{Nmodegraph}), we reject direct, physical phase control and use digital postprocessing to transform the covariance matrix to the desired form on which we then verify its entanglement properties.
However, we do preserve the coherence between pump tones by phase locking so that the relative phases do not fluctuate over time.
By applying a postprocessing phase rotation for each mode separately, we bring the covariance matrix into the symmetric form (see Sect. \ref{tripartite}).

For the analysis of full inseparability of the covariance matrix according to the PPT criterion, we evaluate the minimum symplectic eigenvalues $\min\lbrace\nu_i\rbrace$ of the following mode permutations: $1-234$, $2-134$, $3-124$, $4-123$.
The experimentally obtained symplectic eigenvalues as function of normalized pump amplitude $A$ are displayed in Fig.\,\ref{PPT} alongside with the corresponding predictions given by our numerical simulations. 
The minimum symplectic eigenvalue $\min\lbrace\nu_i\rbrace=0.79\pm0.018$ is reached, while all of the eigenvalues in the normalized pump amplitude range $0.01\lesssim A \lesssim 0.15$ are less than 1.
Compared with the minimum symplectic eigenvalues in Fig.\,\ref{PPT3modeSlice}, we may conclude that the influence of BS correlations on $\min\lbrace\nu_i\rbrace$ value is less in the quadripartite state than for the tripartite case.

The GME criterion for four modes as a function of the normalized pump amplitude is depicted in Fig.\,\ref{PPT}b; the symbols display data while the simulation result is indicated by the solid curve. 
As was discussed in Sect. \ref{theorFound}, the optimized weights in GME inequality Eq.\,(\ref{Scrit_4}) are chosen in the same manner as in the tripartite case: $h_1=g_1=1$ and $h_i = h, g_i = g$, $i=\lbrace{2,3,4\rbrace}$.
The strongest genuine entanglement  $S=0.84 \pm 0.02$ is observed at $A\eqsim 0.08 $ pump amplitude using the weights $h_i=\lbrace1,-0.51,-0.51,-0.51\rbrace$ and $g_i=\lbrace1,0.69,0.69,0.69\rbrace$.
The numerical simulation provides $h$ and $g$ coefficients that coincide with the experimental values with 1\% error, which strongly establishes that the states produced in the experiment coincide with the ones that were obtained and analysed in the numerical model.

The covariance matrices obtained in the experiment and using numerical simulation are presented in Figs.\,\ref{CovMat_exp}a and \ref{CovMat_exp}b, respectively. They are determined at the strongest entanglement point reached at $A\eqsim 0.08 $.
TMS type of correlations are seen in the mode combinations $1-2$, $2-3$, $3-4$ and $1-4$. Subspaces corresponding to BS correlations are visible in the plot as product distributions in $1-3$ and $2-4$ subpartitions. The covariance matrix structures illustrated in Fig.\,\ref{CovMat_exp}  correspond directly to the $\mathscr{\tilde{H}}$-graph structures shown on Fig.\,\ref{Nmodegraph}b.
In general, we conclude that for the employed pump configuration, the genuine quadripartite entanglement appears in the amplitude range $0.01\lesssim A<0.13$.

\section{Discussion}

The control of bisqueezed tripartite and generalized H-graph ($\mathscr{\tilde{H}}$-graph) quadripartite states by relative positioning of the pump frequencies and their phases is indicative of the strong potential of these methods for CV quantum state processing. 
The basic parametric microwave setting allows for enhancement in the number of spectral modes by additional pump tones, which leads to generation of more complex, entangled $\mathscr{\tilde{H}}$-graph states. 
Enhanced number of modes requires larger bandwidth, which calls for broadband parametric devices such as TWPAs \cite{Macklin2015,zorin2019flux,Perelshtein2021,Esposito2021,qiu2022} or broadband JPAs \cite{Roy2015a, Elo2019} in order to avoid problems with spectral mode crowding.

Our approach based on QLE puts in evidence additional correlations, which are captured by the definition of $\mathscr{\tilde{H}}$-graph states. The correlations arise naturally from the connection between intracavity modes and the input vacuum modes, due to which the same vacuum fluctuations may act in the downconversion of more than one quanta. In the literature on cluster and H-graph states, the adjacency matrix for H-graphs is defined via the matrix specified in the multimode squeezing Hamiltonian. The QLE analysis corresponds to the expansion of the multimode squeezing operator up to second order, which leads to the appearance of beamsplitter correlations in the adjacency matrix. In Appendix \ref{analytics} (Eqs. (\ref{intMatrix4modes_NOBS})--(\ref{covarianceEq4modes_NOBS})) we show how to use well-chosen relative pump phase values in the quadripartite case to prepare an entangled square lattice state -- that is, a state without BS correlations. For the case of very large squeezing, $\mathscr{\tilde{H}}$-graph state can be regarded as an approximation of a 4-node cluster state, minimizing errors in gate operations of measurement-based CV quantum computing. 

Cluster states form a promising platform for scalable quantum information processing. 
In \textit{one-way quantum computing} \cite{PhysRevLett.86.5188}, the entire computational resource is provided by the entanglement of the cluster state. The processing is based on quantum measurements which facilitate  gate operations as well as the read-out of the final result. 
However, cluster states can be obtained from graph states only in the mathematical limit of large squeezing parameter \textcolor{blue}{\cite{Zhang2006,Menicucci2006,Gu2009,PhysRevA.102.052424}}.
For quantum information processing steps, it is sufficient to perform sub-cluster measurements in specified order using a suitable computational basis.  
In Refs.\,\onlinecite{Menicucci2006,alexander2016flexible,PhysRevA.90.043841} different computation scenarios based on resources provided by squeezing generators and beamsplitters are described. 
Encoding, gate and measurement operations have been so far considered in optical circuits for continuous variable quantum data and can be efficiently extended to the microwave realm.
In this work, we have have utilized this correspondence between optics and microwaves and demonstrated $\mathscr{\tilde{H}}$-graph state encoding. 

In contrast to computational models for graph states \cite{menicucci2011graphical} considered as \textit{ideal} clusters, hardware based on finite squeezing with noise and decoherence requires error correction  procedures \cite{Menicucci2014,larsen2020architecture, PhysRevApplied.15.034073} to provide reliable CV computation.
Using the presented scheme one can implement error correction codes based on the idea of repetitions of selective measurements and new encoding of $\mathscr{\tilde{H}}$-graph states before each gate operation. In Ref.\,\onlinecite{PhysRevResearch.2.023138} a multidimensional platform for scalable quantum computing has been proposed, based on cluster states created using microring resonators; also multiple frequency combs \cite{PhysRevA.90.043841} created by optical parametric amplifiers and beamsplitters can serve as an excellent platform for quantum computation.
Our work shows that the methods of generation of highly-entangled CV states are not restricted to just optical parametric amplifiers, but the methods can be carried over into the microwave domain by employing   parametric Josephson junction devices for creation of topologically involved and structurally versatile $\mathscr{\tilde{H}}$-graph states.

An implementation of the universal quantum computer based on bosonic modes with the possibility of hardware-efficient quantum error correction \cite{Hillmann2020} requires efficient generation of continuous-variable quantum resources. The genuine entanglement between several bosonic modes could potentially be employed for error-correctable codeword states \cite{gao2019entanglement}.
Besides potential in error correction, the introduction of entanglement into quantum measurement implementations leads to a quantum advantage in the detection process when detection is performed in the presence of high level of noise and loss \cite{lloyd2008enhanced}. 

Increase of the number of entangled spectral modes is essential for future technological application of these CV quantum state generation methods. The limiting factors are the requirements of high precision for the pump frequency and its phase, the stability of the biasing flux, and possible crowding of modes within a narrow-band JPA resonance. However, recently it has been demonstrated that entanglement can be generated in low-loss traveling wave parametric amplifiers  \cite{Perelshtein2021,Esposito2021,qiu2022}. This opens a way to significant increase in the number of entangled modes.

\section{Conclusion}

In this work, we presented a practical scheme for generation of controllable multipartite entanglement from vacuum fluctuations, based on multitone pumping scheme of a JPA, which facilitates pivotal resources for quantum technologies at microwave frequencies.
While optical schemes for multipartite entanglement generation operate on even larger clusters, they lack versatility and are limited to optical frequencies as such. 
On the other hand, our scheme allows for a flexible increase in the number of modes and control of the entanglement configuration among the modes by adjusting pumping on the same device, whereas optical setups call for massive hardware reconfiguration when the entanglement structure is altered.
Through phase and amplitude variation of the microwave pump tones, we reach a comprehensive control over the entanglement structure within the spectral modes of a single JPA cavity mode, which we experimentally verify in detail for the tripartite case. 

Using the developed scheme, we made the first successful demonstration of an on-demand tunable, fully inseparable and phase controllable genuinely entangled tripartite and quadripartite states in a superconducting system.
The presence of multipartite quantum correlations was verified using the covariance matrix formalism and genuine entanglement criteria constructed from the measured quadratures.
Experimental results were accurately reproduced by calculating symplectic eigenvalues of a partially-transposed covariance matrix for full inseparability detection as well as computing GME criteria in normalized pump amplitude in range $0<A<0.5$ ($0<A <0.25$) and verified  genuine entanglement in the range of $0.01\lesssim A<0.4$ ($0.01\lesssim A<0.13$) for the tripartite (quadripartite) state. 

We provided results of phase-dependent GME criterion for bisqueezed state. With optimal phase shift between two pumping tones $\Delta\varphi=-120^\circ$ minimum value of criterion $S=0.75 \pm 0.05$ was obtained. This result were also faithfully reproduced by numerical simulations.

In our analytical derivations, we  demonstrated additional control possibilities over the BS correlations in the covariance matrix of quadripartite $\mathscr{\tilde{H}}$-graph state. To visualize the formed entanglement structure, we provided an extension for the known H-graph adjacency matrix: besides TMS, it includes BS correlations between the vacuum modes. The QLE approach was used to introduce such an adjacency matrix and to connect it to the general approach starting from multimode squeezing operator and the TMS Hamiltonian for the multi-mode case with multiple pumps. 
As shown in Appendix \ref{analytics}, BS correlations can be fully suppressed by implementing a 180$^o$ phase shift of one pump. Such a phase combination creates a distinct square-lattice H-graph state which, for the limit of infinite squeezing parameter, transforms to a square-lattice cluster state.  

Additional TMS correlations can be introduced by inserting new pump tones, which can change the nature of the entangled states drastically.
For example, using two additional pump tones with half frequencies at $\lbrace-\frac{\Delta}{2};\frac{\Delta}{2}\rbrace$, we are able to connect all 4 modes with TMS correlations and thereby achieve a GHZ-like state. Furthermore, by tuning the phases of the pumps, the state can be converted into square-lattice H-graph state.
With the bandwidth improvements provided by the state-of-the-art superconducting parametric devices, such as the broadband, low-loss travelling wave parametric amplifier \cite{Perelshtein2021,Esposito2021,qiu2022}, we expect a substantial increase in the number of entangled modes, which facilitates generation of highly-squeezed square-lattice H-graph states for CV quantum computation at microwave frequencies.

\section*{Acknowledgments}
We thank Mark Dykman, Stefano Pirandola, Olivier Pfister, Aidar Sultanov, Shruti Dogra and Marko Kuzmanović for fruitful discussions and useful comments. This work was supported by the Academy of Finland through the Finnish Center of Excellence in Quantum Technology QTF (projects 312295, 312296, 336810, 336813). 
This project has received funding from the European Union's Horizon 2020 research and innovation programme under grant agreement no. 862644 (FET-Open project QUARTET) and grant agreement no. 824109 (European Microkelvin Platform), as well as ERC grant agreement no. 670743 (QuDeT). GSP and KVP also would like to thank Saab for support, under a research agreement between Saab and Aalto University.
The work of PJH and IL was supported by MATINE grant VN/13774/2020-PLM-49. 

\bibliography{references.bib}

\newpage

\section{Analytical methods, experimental procedures, and entanglement analysis}
\label{methods}
\subsection{Details of theoretical description}
The Hamiltonian of JPA system is given by

\begin{eqnarray}
{\hat{H}}_{\text{sys}}=\hbar\omega_r \hat{a}^\dagger \hat{a} & + \frac{\hbar}{2}\sum_{d=1}^{p}\left[\alpha^{*}_d e^{i\omega_d t}  + \alpha_d e^{-i\omega_d  t}\right](\hat{a}^{2} + \hat{a}^{\dagger 2}) \nonumber \\
& + \hbar K (\hat{a}+ \hat{a}^{\dagger})^{4},
\end{eqnarray}
where $\hat{a}$($\hat{a}^{\dagger}$) is the annihilation (creation) operators for cavity photons, $\alpha_d$ is the complex amplitude for pump tone $d$, and $K$ denotes the strength of the Kerr nonlinearity term. 
Using the average of $p$  pump  tones $\omega_d$, $d=\left\lbrace  1,\dots,p \right\rbrace  $, we define the detuning between the half pump frequency and the resonator frequency: $\Delta_r=\omega_r-\frac{\omega_{\Sigma}}{2}$, $\omega_{\Sigma}=\frac{\sum_{d=1}^{p}\omega_d}{p}$. 

For each of the $p$ pump tones, we define the detuning from the average frequency $\Delta_d=\omega_d-\omega_{\Sigma}, d=\left\lbrace  1,\dots,p \right\rbrace  $. 
By applying the rotating wave approximation in the frame $\omega_\Sigma/2$ ($\tilde{a}(t) = \hat{a}(t)e^{i\omega_\Sigma t/2}$) and leaving only the effective high-order terms, we obtain for the nonlinear part of the Hamiltonian

\begin{eqnarray}
{H}_{\text{sys,rwa}}(t)={\hbar}\Delta_r \tilde{a}^\dagger \tilde{a} & + \frac{\hbar}{2}\sum_{d=1}^{p}(\alpha^{*}_d e^{i\Delta_d t}\tilde{a}^{2} + \alpha_d e^{-i\Delta_d t} \tilde{a}^{\dagger 2}) \nonumber \\
& + 6{\hbar}{K} \tilde{a}^{\dagger}\tilde{a}^{\dagger} \tilde{a} \tilde{a}.
\end{eqnarray}
As usual, the bosonic commutation relationships are valid for the cavity modes
$\left[ \tilde{a},\tilde{a}^{\dagger}\right]=1.$

The parametric resonator is coupled to a transmission line via the signal port and to the thermal bath via a linear dissipation port.
The coupling Hamiltonian associated with the signal port is given by
\begin{equation}
H_{\text{sig}}=\hbar \int d\omega \left( \tilde{b}^{\dagger}\tilde{b} + \kappa \tilde{a}^{\dagger} \tilde{b} - \kappa^{*} \tilde{b}^{\dagger} \tilde{a}\right),
\end{equation}
where creation and annihilation operators $\tilde{b}^{\dagger}$ and $\tilde{b}$ refer to modes in the transmission line, and $\kappa$ denotes the coupling rate. The Hamiltonian related to the linear dissipation port
\begin{equation}
H_{\text{loss}}=\hbar \int d\omega \left( \tilde{c}^{\dagger}\tilde{c} + \gamma \tilde{a}^{\dagger} \tilde{c} - \gamma^{*} \tilde{c}^{\dagger}\tilde{a}\right),
\end{equation}
where $\tilde{c}^{\dagger}$ and $\tilde{c}$ describe creation and annihilation of thermal bath modes and the rate $\gamma$ represent the coupling of cavity modes to the linear dissipation port. The transmission line and bath operators obey the commutation relations
\begin{equation}
\left[ \tilde{b}(\omega),\tilde{b}^{\dagger}(\omega^{\prime})\right]=\left[ \tilde{c}(\omega),\tilde{c}^{\dagger}(\omega^{\prime})\right]=\delta(\omega-\omega^{\prime})
\end{equation}
and
\begin{eqnarray}
\left[ \tilde{b}(\omega),\tilde{c}^{\dagger}(\omega^{\prime})\right]  = & \left[ \tilde{c}(\omega),\tilde{b}^{\dagger}(\omega^{\prime})\right] = \left[ \tilde{b}(\omega),\tilde{b}(\omega^{\prime})\right] \nonumber \\
& =\left[ \tilde{c}(\omega),\tilde{c}(\omega^{\prime})\right]=0.
\end{eqnarray}
The total Hamiltonian can be conveniently written as a sum of the separate parts given above:
\begin{equation}
{H}(t)=H_{\text{sys,rwa}}(t)+H_{\text{sig}}+H_{\text{loss}}
\end{equation}

For further analysis and for our simulations, we use the Quantum Langevin Equation (QLE) for the cavity operator $\tilde{a}(t)$:
\begin{eqnarray}
\label{QLE}
\dot{\tilde{a}}(t)=(-i\Delta_r &- \frac{\kappa+\gamma}{2})\tilde{a} - i \sum_{d=1}^{p}\alpha_d e^{i\Delta_d t} \tilde{a}^\dagger \nonumber \\
& + \sqrt{\kappa}\tilde{b}_{in}+\sqrt{\gamma}\tilde{c}_{in} - 12i{K} \tilde{a}^{\dagger} \tilde{a} \tilde{a}, 
\end{eqnarray}
where the presence of the Kerr term allows us to consider dynamics of the parametric resonator above the critical oscillation threshold. To obtain the modes coming out from the cavity, we employ the standard input-output formalism which yields the relationship:
\begin{equation}
\label{IOE}
\tilde{b}_{out}(t)=\tilde{b}_{in}(t)-\sqrt{\kappa}\tilde{a}(t).
\end{equation}
Eqs. (\ref{QLE}) and (\ref{IOE}) are used in our numerical simulations with Matlab ODE45 solver.

\subsection{Full inseparability}

Assuming that the microwave fields produced by the JPA below the critical pumping threshold are Gaussian \cite{Wilson2018generatingmultimode}, the states with multiple spectral modes can be fully characterized by measuring the covariance matrix of corresponding in-phase $I$ and quadrature $Q$ voltages.  
For the measurement of tripartite correlations, we collect quadrature data for $0.8$ seconds at every phase difference and pump amplitude value, without averaging.
For the quadripartite case, we repeat the experiment 20 times at each pump power value and every quadrature sequence has a duration of $1.3$ seconds. 

The quantum quadratures $ \tilde{x}_i = \frac{\tilde{a}_i+\tilde{a}_i^{\dagger}}{2}$ and $\tilde{p}_i=\frac{\tilde{a}_i-\tilde{a}_i^{\dagger}}{2i}$ can be combined into a 2N-long column vector operator for the N-mode state $\tilde{r} = \left( \tilde{x}_1, \tilde{p}_1, \dots \tilde{x}_N, \tilde{p}_N \right)^{T}$. The commutation relations can be written down in a skew-symmetric, block-diagonal matrix form \cite{Simon2000}:
\begin{equation}
\left[\tilde{r}_i, \tilde{r}_j\right]=\frac{i}{2}\Omega_{ij} \text{ and } 
\mathbf{\Omega} = \bigotimes_{i=1}^N
\begin{bmatrix}
0 & 1\\
-1 & 0
\end{bmatrix}
.
\end{equation} 
The covariance matrix $\mathbf{V}$ is given by elements  $V_{ij}=\frac{1}{2}\left<\Delta\tilde{r}_i \Delta\tilde{r}_j + \Delta\tilde{r}_j \Delta\tilde{r}_i \right> - \left<\Delta\tilde{r}_i \right> \left<\Delta\tilde{r}_j\right>$ where we have defined standard error $\Delta \tilde{r}_i=\tilde{r}_i-\left<\tilde{r}_i\right>$ and $\left<\tilde{r}_i\right>= \text{tr}\left(\tilde{r}_i \hat{\rho}\right)$. The uncertainty principle requires that
\begin{equation}
\mathbf{V}+\frac{i}{4}\mathbf{\Omega}\geq0
\end{equation}
applies for a physical covariance matrix.

For verification of entanglement, we may investigate a modified equation
\begin{equation}
\label{uncertaint}
\mathbf{V}^\prime+\frac{i}{4}\mathbf{\Omega}\geq0,
\end{equation}
where $\mathbf{V}^\prime_k=\pmb{\lambda}_k \mathbf{V} \pmb{\lambda}_k$, $\pmb{\lambda}_{k}$ is diagonal matrix  with ones entries, except of that related to $k$-th mode, with value of $-1$.  For example, transformation with $\pmb{\lambda}_{k\equiv N}=\text{diag}(1,1,...,1,-1)$ means a partial transposition of the covariance matrix with respect to the last mode.
The Positive Partial Transpose (PPT) criterion for multipartite case requires that there is a violation of Eq. (\ref{uncertaint}) when applying a partial transposition with respect to each from full set of modes: $\mathbf{V}^\prime_k \geq \frac{i}{4}\mathbf{\Omega}$.  In Ref. \onlinecite{Loock2003Detecting}, the entangled states are classified in accordance to the number of modes for which the condition (\ref{uncertaint}) is broken. We follow this approach to demonstrate the highest class - \textit{full inseparability} -  in four mode case. 

Unitary operations which retain the Gaussian character of the states, e.g. squeezing, are of particular importance. Such operations on the Hilbert space correspond to a linear transformation $P$ in the phase-space which preserve the symplectic form, i.e.,
\begin{equation}
\mathbf{\Omega} = \mathbf{P}^T \mathbf{\Omega} \mathbf{P}.
\end{equation}
Symplectic transformations on a 2N-dimensional phase-space form the real symplectic group denoted as $ Sp(2N; \textbf{R})$, which is a proper subgroup of the special linear group of 2N $\times$ 2N matrices \cite{SimonQuantum-noiseMatrix}. By utilizing Williamson’s theorem \cite{Williamson1936Algebraic}, any covariance matrix can be expressed in the Williamson normal form:

\begin{equation}
\mathbf{\tilde{V}}_k= \mathbf{P}^T \mathbf{V}^\prime_k \mathbf{P}
\end{equation}
where $\mathbf{\tilde{V}}_k$ is a 2N-dimensional diagonal matrix consisting of the symplectic eigenvalues, $\tilde{\nu}_k$, of the covariance matrix. The symplectic eigenvalues are called the
symplectic spectrum which provides a practical means to verify physicality and various entanglement criteria. Separability is in force, when condition $\tilde{\nu}_k \geq 1/4$ fulfilled for $\mathbf{\tilde{V}}_k$.

For convenience, we insert an additional factor of $4$ to the covariance matrix and work with fluctuations with zero mean values: ${V}^*_{ij}=2\left<\Delta\tilde{r}_i \Delta\tilde{r}_j + \Delta\tilde{r}_j \Delta\tilde{r}_i \right>$. Consequently, for evidence of 'fully inseparable' states, we need to find minimum symplectic eigenvalues with $\tilde{\nu}_k^* < 1$ for each partial transposition $k$.

\subsection{System gain calibration} \label{systgain}
\begin{figure}
	
	\centering
	\includegraphics[ width=0.8\linewidth]{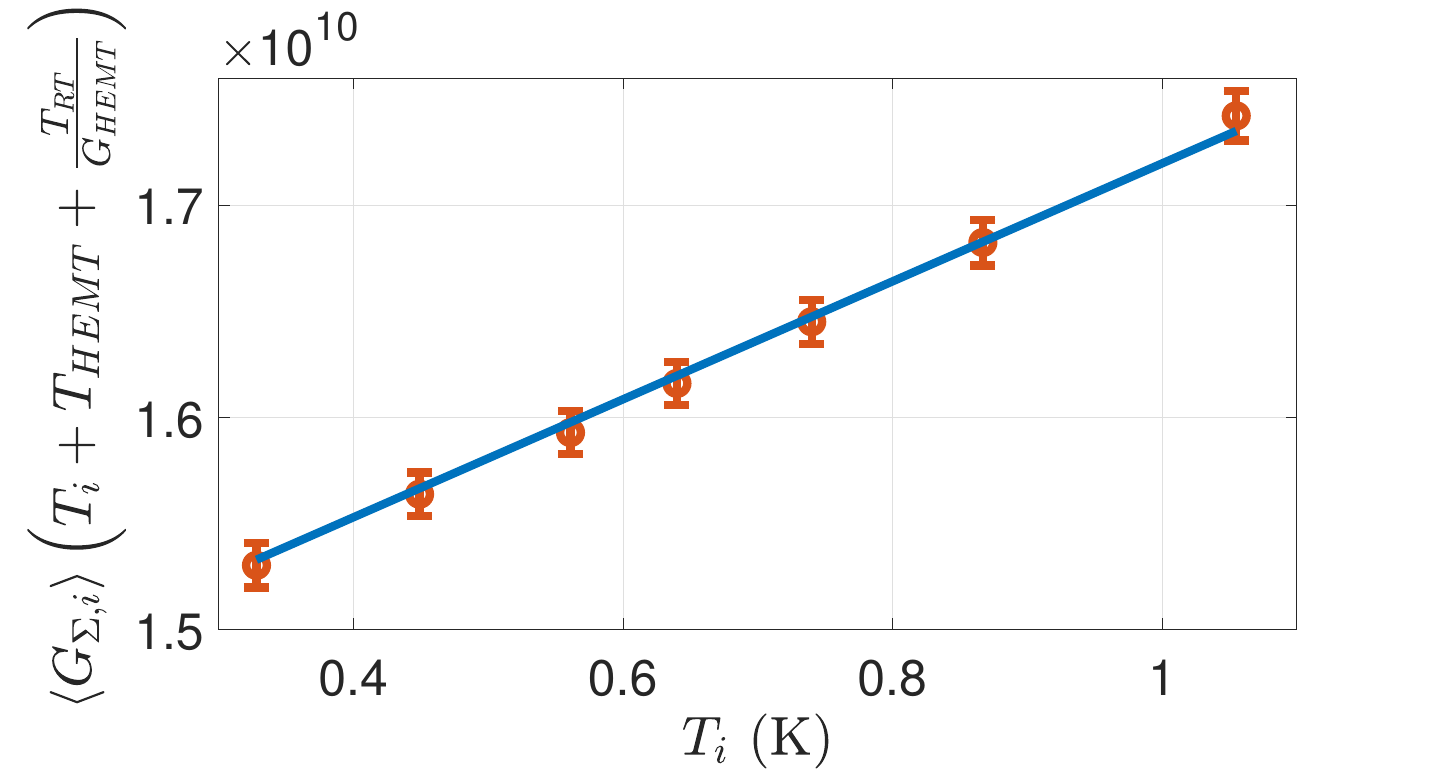}
    	\caption{Gain calibration using linear temperature dependence of the measured thermal noise spectral density of a 50 Ohm terminator measured as a function of $T_s$, the source temperature. The average total  gain $\left<G_{\Sigma, i}\right>=94.4 \pm 0.2$ dB over the cavity resonance is obtained from the  linear fit (in red) to the data. This gain value $\left<G_{\Sigma, i}\right>$ also includes frequency mixing losses/amplification in the signal analyzer circuit part. 
    	The term $T_{\text{preamp}}=T_{\text{HEMT}}+\frac{T_\text{RT}}{G_\text{HEMT}}=5.2\pm0.25$K characterizes the equivalent noise temperature of the amplifiers; the largest contribution originates from the cooled HEMT amplifier at 4 K. The value $\coth(\frac{hf_i}{2 k_b T_\text{preamp}})$ sets the background for the diagonal elements in the covariance matrix $4\mathbf{V}_\text{off}$.}
	\label{noise_cal}
\end{figure}

Our system gain calibration procedure consists of a measurement of Johnson-Nyquist noise spectral density emitted by a 50 $\Omega$ termination at different temperatures. Assuming perfect matching of the source and load impedances, the received power per unit of bandwidth can be written by applying the Friis formula: the measured noise is given by the noise temperature of the source $T_s$, the contribution of the cooled amplifier $T_{\text{HEMT}}$, and the noise of the room-temperature amplifiers $T_{\text{RT}}$ multiplied by the system gain $G_{\Sigma, i}=G_{\text{HEMT}_i}G_{\text{RT}_i}$:
\begin{equation}
\frac{\left< I_i^2 + Q_i^2 \right>}{ Z_0 \Delta f_i } = k_b G_{\Sigma,i} \left( T_i+T_{\text{HEMT}}+\frac{T_{\text{RT}}}{G_{\text{HEMT}}}\right).
\label{gain_cal}
\end{equation}
Here $i$ refers to the frequency of the spectral mode and $\Delta f_i $ refers to the bandwidth of the detection of quadratures $I_i$ and $Q_i$. The total gain $G_{\Sigma, i}$ was separately determined for different spectral modes.

Fig. \ref{noise_cal} displays the measured noise power per unit band as a function of sample temperature $T_s$, averaged over frequencies covering the resonance curve. By fitting a line to the data, we obtain $\left<G_{\Sigma, i}\right>=94.4 \pm 0.2$ dB for the average total gain. The linear fit in Fig. \ref{noise_cal} is performed at $T>0.2$ K, which allows us to neglect the corrections from the $\coth(\hbar \omega/2k_b T_s)$.

The error in the system gain calibration results in  uncertainty in the symplectic eigenvalues on the order of 2\%, i.e. the eigenvalues fall in the range of  $\min\lbrace\tilde{\nu}_k^*\rbrace$=$\min\lbrace\tilde{\nu}_k^*\rbrace \pm0.018$ for each partial bipartition. Random variations of the system parameters were reduced by averaging the outcome by ten to twenty times.

\subsection{System parameter fitting} \label{systpara}

\begin{figure}
	\subfloat[1\linewidth][]{
		\adjincludegraphics[ width=.85\linewidth]{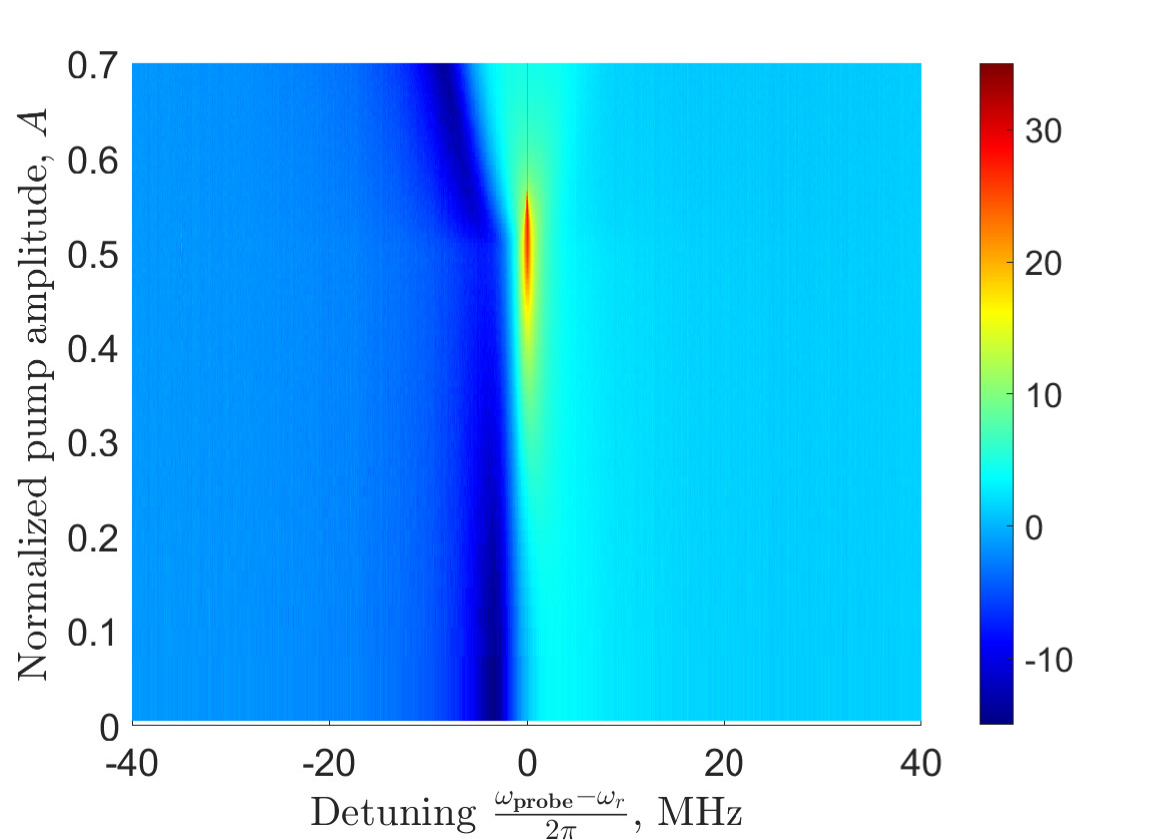}}\\
	\subfloat[1\linewidth][]{
		\adjincludegraphics[ width=.85\linewidth]{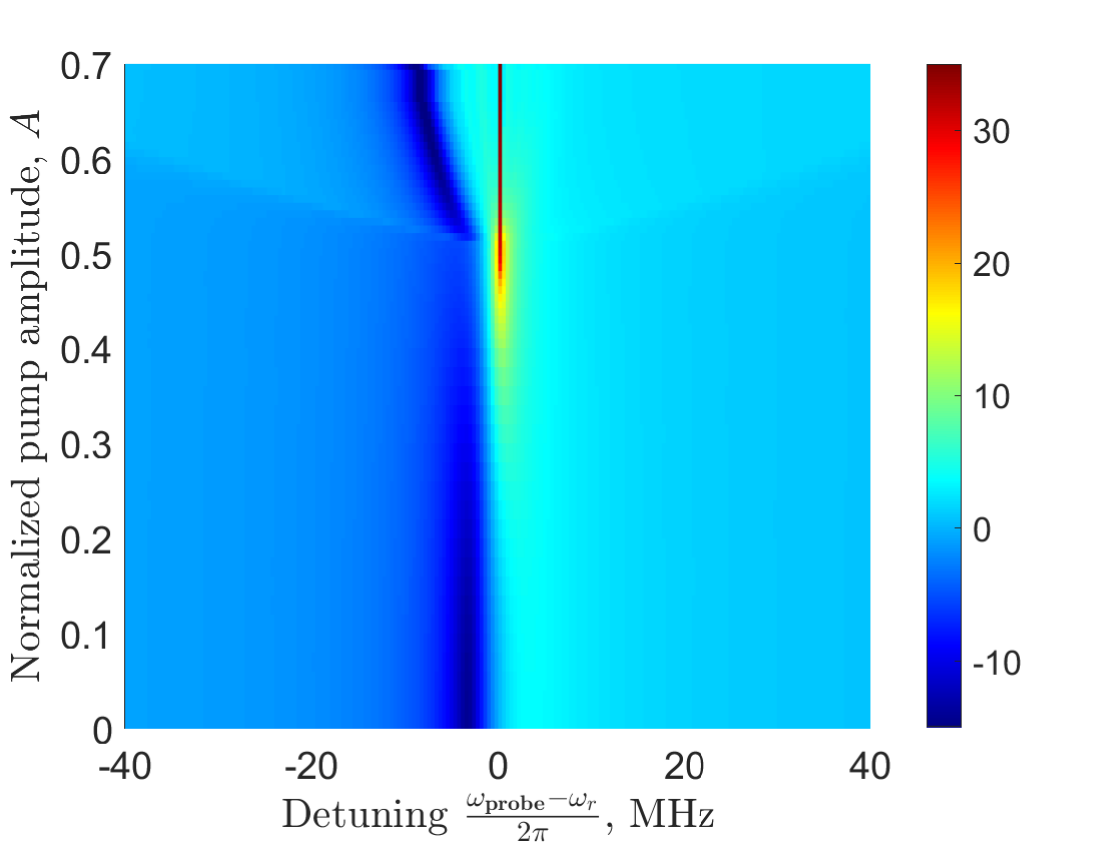}
		}
	\caption{Plots (a) and (b) show gain coefficient measurement and QLE's simulation results (Eq. (\ref{QLE})) which used for fitting coupling and Kerr constants. Vertical axis represents normalized pump amplitude $A=\frac{\alpha}{\kappa+\gamma}$. Horizontal axis represents detuning between probe signal frequency and resonance frequency $\frac{\omega_\text{probe} - \omega_r}{2\pi}$. Pumping carried out in degenerate mode, $\omega_d=2\omega_r$. Fano resonance picture, given in experimental plot, explained by phase shift between the cavity and input modes and described by complex rate value of $\kappa$.     
	}
	\label{gain_measurement_and_simulation}
\end{figure}

In order to determine coupling rates $\gamma$ and $\kappa$ introduced in Section \ref{theorFound}A, we characterized our nonlinear resonator as a two-port device using a vector network analyzer.
For the characterization, we chose the optimal DC operating point $\Phi_{DC} = 0.383\,\Phi_0$ depicted in Fig.\,\ref{expFig}b.
At this DC-flux, we measured the resonance curve in the absence of the pump in order to estimate the external and internal loss rates $\kappa$ and $\gamma$, respectively.
By fitting the measured resonance curve to the analytical solution of the QLE ($\tilde{b}_{out}(\omega)/\tilde{b}_{in}(\omega)$), derived for the linear case without any pump drive, we obtain the coupling coefficients $\frac{\kappa}{2\pi} = 4.44$ MHz and  $\frac{\gamma}{2\pi}=2.30$ MHz. 
The employed analytical solution, displayed in Eq. (\ref{fitting}), was derived from the full QLE in Eq. (\ref{QLE}) without taking the nonlinear part $-iK \tilde{a}^{\dagger} \tilde{a} \tilde{a} $ into account:
\begin{equation}
	\label{fitting}
	\frac{\tilde{b}_{out}(\omega)}{\tilde{b}_{in}(\omega)}=1-\frac{\kappa}{(-i(\omega-\omega_r) +\frac{\kappa+\gamma}{2})}.
\end{equation}
For fitting of the Kerr constant $K$, we employed the whole form of the QLE in the rotating wave approximation (\ref{QLE}). 
By comparing the measured and simulated gain coefficients $G(\omega_\text{probe}-\omega_r,A)$ (Fig. \ref{gain_measurement_and_simulation}) in the cavity at large pump amplitudes, we obtain an estimate  $K = 6.5\omega_r$ for the Kerr constant.

\subsection{Cavity phase response}
\label{phase_resp}
\begin{figure}
	
	\centering
	\includegraphics[ width=0.8\linewidth]{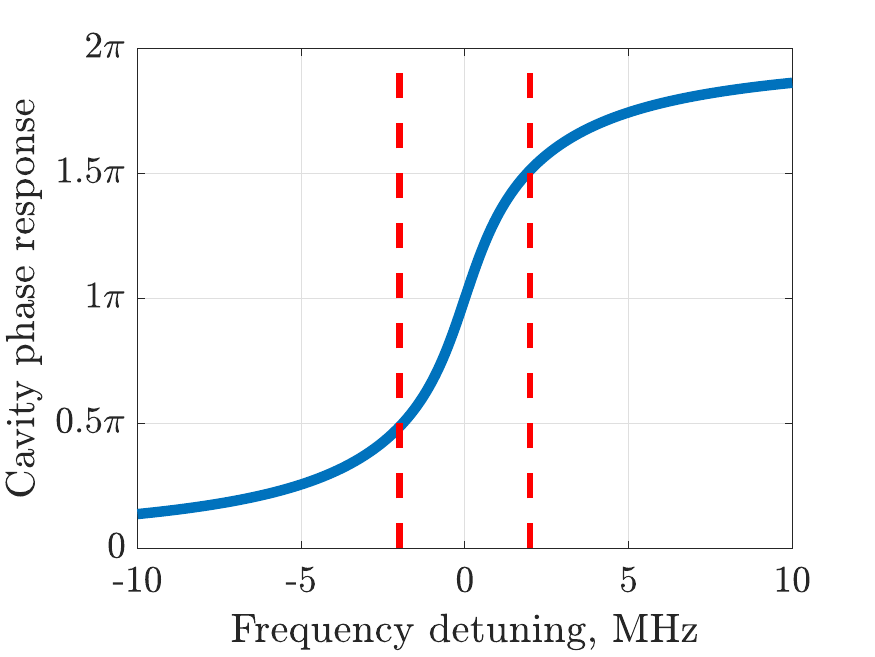}
    	\caption{Cavity phase response given by fitted experimental parameters $\frac{\kappa}{2\pi} = 4.44$ MHz and  $\frac{\gamma}{2\pi}=2.30$ MHz. Vertical dash lines show center frequencies of first and last modes. Corresponding phase shifts applied to pump tones to reach "symmetric" covariance matrix view on double frequencies are $\Delta\varphi_1=\frac{\pi}{2}-\frac{\pi}{4}=\frac{\pi}{4}$ and $\Delta\varphi_2=\frac{\pi}{2}+\frac{\pi}{4}=\frac{3\pi}{4}$ with corresponding phase shift between pump tones $\frac{\pi}{2}$, given in results (See. Fig. \ref{3modes}). Half of applied phase shift described by pump tones on double resonance frequencies. Additional phase shift with increase of $A$ relates to modification of phase response curve during pumping.}
	\label{cav_phase} 
\end{figure}

Optimal value of GME criterion, which governed by "symmetric" covariance matrix view in tripartite mode case, can be obtained with $\lbrace\frac{\pi}{2};\frac{\pi}{2}\rbrace$ only if modes reshuffling suffers no additional phase rotations (See next Subsection). However, in experiments we deal with finite values of coupling and dissipation loss rates. Cavity phase response becomes crucial figure in pump tones phase shift adjustments. Cavity phase response illustrated on Fig. \ref{cav_phase}. 

\subsection{Multifrequency correlations in terms of QLE with 3 and 4 spectral modes}

\label{analytics}
As discussed in Section \ref{theorFound}C, our measurement setting probes outgoing waves from the parametric resonator, which brings about slight differences with standard quantum optics schemes where the entanglement analysis is based on the Hamiltonian of the system. In our case, the QLE provides a good description, and here we derive the relevant matrix equations describing the coupling of the different outgoing spectral modes under two and three pump tones (3 and 4 spectral modes, respectively).

\textbf{3 mode case.}
Let us define $\tilde{a}$ as a vector of spectral modes:
\begin{equation}
    \tilde{a}=\lbrace{\tilde{a}_1,\tilde{a}_2,\tilde{a}_3,\tilde{a}^\dagger_1,\tilde{a}^\dagger_2,\tilde{a}^\dagger_3\rbrace}^{T};
  \end{equation}  
  where the creation $\tilde{a}_i^{\dagger}=\tilde{a}_i^{\dagger}(t)$ and annihilation $\tilde{a}_i=\tilde{a}_i(t)$ operators are time-dependent.  After Fourier transform, 
  \begin{equation}
    \tilde{a}(\omega)=\lbrace{\tilde{a}_1(\omega),\tilde{a}_2(\omega),\tilde{a}_3(\omega),\tilde{a}^\dagger_1(-\omega),\tilde{a}^\dagger_2(-\omega),\tilde{a}^\dagger_3(-\omega)\rbrace}^{T}.
  \end{equation}  
  We define our spectral modes $a_i$ as $\lbrace(-\frac{3\Delta}{2},-\frac{\Delta}{2});(-\frac{\Delta}{2},\frac{\Delta}{2});(\frac{\Delta}{2},\frac{3\Delta}{2})\rbrace$ according pump tone positions $\lbrace-\Delta,\Delta\rbrace$ (see Fig. \ref{modes_separation_fig}). 
  Similarly, we define for the input and output modes $\hat{b}_{\text{in}/\text{out}}$:
  \begin{eqnarray}
    \tilde{b}_{\text{in/out}}=& \lbrace{\tilde{b}_{\text{in}1/\text{out}1},\tilde{b}_{\text{in}2/\text{out}2},\tilde{b}_{\text{in}3/\text{out}3}}, \nonumber \\
  &  \tilde{b}^\dagger_{\text{in}2/\text{out}2},\tilde{b}^\dagger_{\text{in}1/\text{out}1},\tilde{b}^\dagger_{\text{in}3/\text{out}3}\rbrace^{T}.
\end{eqnarray}
The commutation relationships for the case of $N$ modes can be conveniently expressed in matrix form. We use the common convention for $\left[\tilde{a}_i,\tilde{a}_j \right]$ from Ref. \onlinecite{SimonQuantum-noiseMatrix}. 

The effect of Kerr nonlinearity is significant only at large pump amplitudes. Hence, we may take the QLE (\ref{QLE}) without the nonlinear part for our treatment.
In theoretical analysis we assume, that spectral modes lay down deep in cavity mode, such that $\Delta \ll \kappa$;  we also neglect internal dissipation expressed by loss rate $\gamma$. For that case phase shift between modes, provided by phase response of the cavity, can be neglected.  
Guided by standard Fourier transform technique for solving linear QLE \cite{yamamoto2016parametric}, we denote $\tilde{a}_i(\omega)=\int \tilde{a}_i(t) e^{i\omega t}dt$ and Fourier transform the QLE term by term. Owing to detuning of the pump tones in the rotating frame, there will be coupling of spectral modes and we have mode index exchange. For example, for $\tilde{a}^\dagger_{1,2}$: $\int \tilde{a}^\dagger_1(t)e^{i\omega t}e^{i\Delta_1 t}dt=\tilde{a}^{\dagger}_2(-\omega)$; $\int \tilde{a}^\dagger_2(t)e^{i\omega t}e^{i\Delta_2 t}dt=\tilde{a}^{\dagger}_1(-\omega)$, while for $\tilde{a}^\dagger_{2,3}$: $\int \tilde{a}^\dagger_2(t)e^{i\omega t}e^{-i\Delta_2 t}dt=\tilde{a}^{\dagger}_3(-\omega)$; $\int \tilde{a}^\dagger_3(t)e^{i\omega t}e^{-i\Delta_3 t}dt=\tilde{a}^{\dagger}_2(-\omega)$.
Thus, it is seen that each pump creates a two-mode squeezed state (TMS) between two neighboring spectral modes independently from RWA's zero-frequency position. 

After Fourier transforming,
\begin{eqnarray}
(-i(\omega-\Delta_r)+&\frac{\kappa}{2})\tilde{a}(\omega)+i\alpha(\int \tilde{a}^\dagger(t)e^{i\omega t}e^{-i\Delta_d t}dt+ \nonumber \\ 
&\int \tilde{a}^\dagger(t)e^{i\omega t}e^{i\Delta_d t}dt)=\sqrt{\kappa}\tilde{b}_{in}(\omega),
\end{eqnarray}
the QLE yields the following system of linear equations:
\small
\begin{eqnarray}
\sqrt{\kappa}\tilde{b}_{\text{in}1}(\omega)=(-i(\omega-\Delta_r)+\frac{\kappa}{2})\tilde{a}_1(\omega)+i\alpha \tilde{a}^\dagger_2(-\omega) \nonumber \\
\sqrt{\kappa}\tilde{b}_{\text{in}2}(\omega)=(-i(\omega-\Delta_r)+\frac{\kappa}{2})\tilde{a}_2(\omega)+i\alpha(\tilde{a}^\dagger_1(-\omega)+\tilde{a}^\dagger_3(-\omega)) \nonumber \\
\sqrt{\kappa}\tilde{b}_{\text{in}3}(\omega)=(-i(\omega-\Delta_r)+\frac{\kappa}{2})\tilde{a}_3(\omega)+i\alpha \tilde{a}^\dagger_2(-\omega) \nonumber \\
\sqrt{\kappa}\tilde{b}^\dagger_{\text{in}1}(-\omega)=(-i(\omega+\Delta_r)+\frac{\kappa}{2})\tilde{a}^\dagger_1(-\omega)-i\alpha^\dagger \tilde{a}_2(\omega) \nonumber \\
\sqrt{\kappa}\tilde{b}^\dagger_{\text{in}2}(-\omega)=(-i(\omega+\Delta_r)+\frac{\kappa}{2})\tilde{a}^\dagger_2(-\omega)-i\alpha^\dagger(\tilde{a}_1(\omega)+\tilde{a}_3(\omega)) \nonumber \\
\sqrt{\kappa}\tilde{b}^\dagger_{\text{in}3}(-\omega)=(-i(\omega+\Delta_r)+\frac{\kappa}{2})\tilde{a}^\dagger_3(-\omega)-i\alpha^\dagger \tilde{a}_2(\omega). \nonumber \\
\label{QLE3modes}
\end{eqnarray}
\normalsize
We cast Eq. (\ref{QLE3modes})  into matrix form:
\begin{equation}
    \mathbf{M}\tilde{a}(\omega)=\sqrt{\kappa}\tilde{b}_{\text{in}}(\omega)
\end{equation}
\begin{equation}
    \mathbf{M}=\begin{bmatrix}
c_1 & 0 & 0 & 0 & i\alpha & 0\\
0 & c_1 & 0 & i\alpha & 0 & i\alpha\\
0 & 0 & c_1 & 0 & i\alpha & 0\\
0 & -i\alpha^\dagger & 0 & c_2 & 0 & 0\\
-i\alpha^\dagger & 0 & -i\alpha^\dagger & 0 & c_2 & 0\\
0 & -i\alpha^\dagger & 0 & 0 & 0 & c_2
\end{bmatrix}
\end{equation}
where $c_1=-i(\omega-\Delta_r)+\frac{\kappa}{2}$ and $c_2=-i(\omega+\Delta_r)+\frac{\kappa}{2}$
Solving for the inverse of $\hat{M}$ and using Eq. (\ref{IOE}), we obtain 
\begin{equation}
    \tilde{b}_{\text{out}}(\omega)=(\mathbf{I}-\kappa\mathbf{M}^{-1})\tilde{b}_{\text{in}}(\omega)
\end{equation}
for the outgoing radiation $\tilde{b}_{\text{out}}(\omega)$ in terms of incoming waves $\tilde{b}_{\text{in}}(\omega)$. 

Because our goal is to determine the structure of the experimental covariance matrix, it is unsatisfactory to consider cavity modes $\tilde{a}$ with equation $\tilde{a}(\omega)=\sqrt{\kappa}\mathbf{M}^{-1}\tilde{b}_{\text{in}}(\omega)$ though it has a more compact final form. However, the presence of the identity matrix $\mathbf{I}$ and the multiplication factor $\kappa$ do not change the final structure. 

Assuming that the pump amplitude $\alpha$ is a real number and $c_1=c_2=c$ (zero detuning case), we have 
\begin{eqnarray}
    \mathbf{M}^{-1}= \frac{1}{c^2-2\alpha^2} \nonumber  \cdot\\
    \begin{bmatrix}
c-\frac{\alpha^2}{c} & 0 & \mathbf{\frac{\pmb{\alpha}^2}{c}} & 0 & -i\alpha & 0\\
0 & c & 0 & -i\alpha & 0 & -i\alpha\\
\mathbf{\frac{\pmb{\alpha}^2}{c}} & 0 & c-\frac{\alpha^2}{c} & 0 & -i\alpha & 0\\
0 & i\alpha & 0 & c-\frac{\alpha^2}{c} & 0 & \mathbf{\frac{\pmb{\alpha}^2}{c}}\\
i\alpha & 0 & i\alpha & 0 & c & 0\\
0 & i\alpha & 0 & \mathbf{\frac{\pmb{\alpha}^2}{c}} & 0 & c-\frac{\alpha^2}{c}
 \end{bmatrix}\label{M1matrix}
\end{eqnarray}
This allows us to draw the generalized $\mathscr{\tilde{H}}$-graph for describing the parametric interaction between the spectral modes, Fig. \ref{Nmodegraph}. The off-diagonal beamsplitter elements proportional to $\alpha^2$ are set in bold in Eq. (\ref{M1matrix}).

Still, we want to  construct the parametric interaction matrix $\mathbf{S}^{-1}$ for quadrature vector operator  $\tilde{r}$. 
Using a linear operator matrix $\mathbf{K}$ to implement a change of basis
\begin{eqnarray}
    \mathbf{K}= \frac{1}{2}
    \begin{bmatrix}
1 & 0 & 0 & 1 & 0 & 0\\
-i & 0 & 0 & i & 0 & 0\\
0 & 1 & 0 & 0 & 1 & 0\\
0 & -i & 0 & 0 & i & 0 \\
0 & 0 & 1 & 0 & 0 & 1\\
0 & 0 & -i & 0 & 0 & i
 \end{bmatrix},
\end{eqnarray}
we obtain by a canonical transformation  $\mathbf{S}^{-1}=\sqrt{\kappa}\mathbf{K}\mathbf{M}^{-1}\mathbf{K}^{-1}$:
\begin{eqnarray}
    \mathbf{S}^{-1}= \frac{\sqrt{\kappa}}{c^2-2\alpha^2} \nonumber \cdot\\
    \begin{bmatrix}
c+\frac{\alpha^2}{c} & 0 & 0 & -\alpha &  \mathbf{\frac{\pmb{\alpha}^2}{c}} & 0 \\
0 & c+\frac{\alpha^2}{c} & -\alpha & 0 & 0 & \mathbf{\frac{\pmb{\alpha}^2}{c}}\\
0 & -\alpha & c & 0 & 0 & -\alpha\\
-\alpha & 0 & 0 & c & -\alpha & 0 \\
\mathbf{\frac{\pmb{\alpha}^2}{c}} & 0 & 0 & -\alpha & c+\frac{\alpha^2}{c} & 0\\
0 & \mathbf{\frac{\pmb{\alpha}^2}{c}} & -\alpha & 0 & 0 & c+\frac{\alpha^2}{c}
 \end{bmatrix}.
\end{eqnarray}
Note, that here the overall structure of the matrix has changed because of the basis change from ladder to quadrature operators. 
This is seen, for example, in the distribution of the off-diagonal beamsplitter correlations (shown in bold).

Since the environment of the cavity is in the ground state, $\tilde{b}_\text{in}$ has a Gaussian covariance matrix of the form $\mathbf{V}_\text{in}=\frac{1}{4}\mathbf{I}$. Consequently, the covariance matrix of the cavity spectral modes $\tilde{a}_i$ can be represented as
\begin{equation}\label{DEF_Va}
    \mathbf{V}_\text{a}=\mathbf{S}^{-1}\mathbf{V}_\text{in}(\mathbf{S}^{-1})^{T}
\end{equation} 
or, equivalently, for output modes $\tilde{b}_\text{out}$: $\mathbf{V}_\text{out}= (\mathbf{I}-\sqrt{\kappa}\mathbf{S}^{-1})\mathbf{V}_\text{in}(\mathbf{I}-\sqrt{\kappa}\mathbf{S}^{-1})^{T}$.
Both forms $\mathbf{V}_\text{a}$ and $\mathbf{V}_\text{out}$ can be employed for studying the structure of parametric interactions between the quadratures, because input-output relationship doesn't change the general structure of the couplings between the quadratures (see below).

As shown in Section \ref{entanglement_results} experimentally, phase shift between pumps changes the appearance of the covariance matrix (see Fig. \ref{3_modes_result}) as well as the strength of genuine multipartite entanglement. A change in the matrix $\mathbf{M}$ due to a phase shift is illustrated in Eq. (\ref{intM3modes}), in which the phase of the first pump has been rotated by $e^{\frac{i\pi}{2}}$. 
\begin{equation}
    \hat{M}=\begin{bmatrix}
c & 0 & 0 & 0 & -\pmb{\alpha} & 0\\
0 & c & 0 & -\pmb{\alpha} & 0 & i\alpha\\
0 & 0 & c & 0 & i\alpha & 0\\
0 & -\pmb{\alpha} & 0 & c & 0 & 0\\
-\pmb{\alpha} & 0 & -i\alpha & 0 & c & 0\\
0 & -i\alpha & 0 & 0 & 0 & c
\end{bmatrix}.
\label{intM3modes}
\end{equation}
The elements affected by the rotation are indicated in bold in the matrix. The elements in bold face indicate coupling between modes $\tilde{a}_1(\omega)\leftrightarrow\tilde{a}_2(\omega)$ while the other off-diagonal elements indicate squeezing across $\tilde{a}_2(\omega)\leftrightarrow\tilde{a}_3(\omega)$.
Note, that phase rotation operates in opposite direction on rows related to $\tilde{b}_{\text{in}}(\omega)$ and $\tilde{b}_{\text{in}}^{\dagger}(\omega)$.

The inversion of the rotated matrix $\mathbf{M}$ yields for the parametric interaction matrix, where all the beamsplitter elements (in bold) have acquired a $\pi/2$ phase shift. This phase shift can be unwound by a phase shift on the second pump, which indicates different phase dependence of the beamsplitter correlations compared with the TMS correlations.
The structure of matrix $\mathbf{M}^{-1}$ in Eq. (\ref{rotatedInvM}) shows that \textbf{phase rotation of specified pump tones does not change parametric interaction form between modes, preserving structure of a bisqueezed tripartite state}. However, as shown in the main text, the  criterion describing the strength of GME (see Eq. (\ref{Scrit_3})) depends on the difference of pump phases and strong genuine entanglement is reached only at specific phase settings.

\begin{eqnarray}
    \mathbf{M}^{-1}= \frac{1}{c^2-2\alpha^2}\cdot \nonumber \\
    \begin{bmatrix}
c-\frac{\alpha^2}{c} & 0 & \mathbf{\frac{\pmb{i\alpha}^2}{c}} & 0 & \alpha & 0\\
0 & c & 0 & \alpha & 0 & -i\alpha\\
-\mathbf{\frac{\pmb{i\alpha}^2}{c}} & 0 & c-\frac{\alpha^2}{c} & 0 & -i\alpha & 0\\
0 & \alpha & 0 & c-\frac{\alpha^2}{c} & 0 & -\mathbf{\frac{\pmb{i\alpha}^2}{c}}\\ 
\alpha & 0 & i\alpha & 0 & c & 0\\
0 & i\alpha & 0 & \mathbf{\frac{\pmb{i\alpha}^2}{c}} & 0 & c-\frac{\alpha^2}{c}
 \end{bmatrix}\label{rotatedInvM}
\end{eqnarray}

The covariance matrix $\mathbf{V}_\text{a}$ obtained from matrix $\mathbf{M}$ in Eq. (\ref{M1matrix}) with zero pump phase shifts is given in Eq. (\ref{covarianceEq3modes0phase}). The corresponding covariance matrix for $\pi/2$ phase rotation in the first pump is displayed in Eq. (\ref{covarianceEq3modes90phase}). The matrix $\mathbf{V}_\text{a}$ in Eq. (\ref{covarianceEq3modes90phase}) has one rotated subspace, corresponding to two quadrature pairs; these rotated components are indicated by bold face. Based on these analytical relationships we conclude that \textbf{control over desired covariance matrix TMS-subspace can be provided by phase rotation of corresponding pump tone}.
Finally, we introduce the same phase rotation $e^{\frac{i\pi}{2}}$ to the second pump. This brings the covariance matrix for the spectral cavity modes
to the "standard-symmetric" form displayed in Eq. (\ref{covarianceEq3modes9090phase}). By comparing Eq. (\ref{covarianceEq3modes0phase}) and Eq. (\ref{covarianceEq3modes9090phase}) we note that the beamsplitter elements (in bold) in the covariance matrix are unchanged (the phase difference between the pumps is the same) while the TMS elements are different.

\small
\begin{widetext}
\begin{equation}
    \mathbf{V}_\text{a}= \frac{\kappa}{4(c^2-2\alpha^2)^2}\cdot\\
    \begin{bmatrix}
 -\alpha ^2+\frac{2 \alpha ^4}{c^2}+c^2 & 0 & 0 & -2 \alpha  c & \pmb{3 \alpha ^2-\frac{2 \alpha ^4}{c^2}} & 0 \\
 0 & -\alpha ^2+\frac{2 \alpha ^4}{c^2}+c^2 & -2 \alpha  c & 0 & 0 & \pmb{3 \alpha ^2-\frac{2 \alpha ^4}{c^2}} \\
 0 & -2 \alpha  c & 2 \alpha ^2+c^2 & 0 & 0 & -2 \alpha  c \\
 -2 \alpha  c & 0 & 0 & 2 \alpha ^2+c^2 & -2 \alpha  c & 0 \\
 \pmb{3 \alpha ^2-\frac{2 \alpha ^4}{c^2}} & 0 & 0 & -2 \alpha  c & -\alpha ^2+\frac{2 \alpha ^4}{c^2}+c^2 & 0 \\
 0 & \pmb{3 \alpha ^2-\frac{2 \alpha ^4}{c^2}} & -2 \alpha  c & 0 & 0 & -\alpha ^2+\frac{2 \alpha ^4}{c^2}+c^2 \\
 \end{bmatrix}
 \label{covarianceEq3modes0phase}
\end{equation}
\end{widetext}
\normalsize

\small
\begin{widetext}
\begin{equation}
    \mathbf{V}_\text{a}= \frac{\kappa}{4(c^2-2\alpha^2)^2}\cdot\\
    \begin{bmatrix}
 -\alpha ^2+\frac{2 \alpha ^4}{c^2}+c^2 & 0 & \pmb{2 \alpha c} & 0 & 0 & \pmb{-3 \alpha ^2+\frac{2 \alpha ^4}{c^2}} \\
 0 & -\alpha ^2+\frac{2 \alpha ^4}{c^2}+c^2 & 0 & \pmb{-2 \alpha c}   & \pmb{3 \alpha ^2-\frac{2 \alpha ^4}{c^2}} & 0  \\
\pmb{2 \alpha  c} & 0& 2 \alpha ^2+c^2 & 0 & 0 & -2 \alpha  c \\
 0 & \pmb{-2 \alpha  c} &  0 & 2 \alpha ^2+c^2 & -2 \alpha  c & 0 \\
0 & \pmb{3 \alpha ^2-\frac{2 \alpha ^4}{c^2}} &  0 & -2 \alpha  c & -\alpha ^2+\frac{2 \alpha ^4}{c^2}+c^2 & 0 \\
 \pmb{-3 \alpha ^2+\frac{2 \alpha ^4}{c^2}} &  0 &-2 \alpha  c & 0 & 0 & -\alpha ^2+\frac{2 \alpha ^4}{c^2}+c^2 \\
 \end{bmatrix}
  \label{covarianceEq3modes90phase}
\end{equation}
\end{widetext}
\normalsize

\small
\begin{widetext}
\begin{equation}
    \mathbf{V}_\text{a}= \frac{\kappa}{4(c^2-2\alpha^2)^2}\cdot\\
    \begin{bmatrix}
 -\alpha ^2+\frac{2 \alpha ^4}{c^2}+c^2 & 0 & 2 \alpha  c & 0 & \pmb{3 \alpha ^2-\frac{2 \alpha ^4}{c^2}} & 0 \\
 0 & -\alpha ^2+\frac{2 \alpha ^4}{c^2}+c^2 & 0 & -2 \alpha  c & 0 & \pmb{3 \alpha ^2-\frac{2 \alpha ^4}{c^2}} \\
 2 \alpha  c & 0 & 2 \alpha ^2+c^2 & 0 & 2 \alpha  c & 0 \\
 0 & -2 \alpha  c & 0 & 2 \alpha ^2+c^2 & 0 & -2 \alpha  c \\
 \pmb{3 \alpha ^2-\frac{2 \alpha ^4}{c^2}} & 0 & 2 \alpha  c & 0 & -\alpha ^2+\frac{2 \alpha ^4}{c^2}+c^2 & 0 \\
 0 & \pmb{3 \alpha ^2-\frac{2 \alpha ^4}{c^2}} & 0  & -2 \alpha  c & 0 & -\alpha ^2+\frac{2 \alpha ^4}{c^2}+c^2 \\
 \end{bmatrix}.
  \label{covarianceEq3modes9090phase}
\end{equation}
\end{widetext}
\normalsize

\textbf{4 mode case.} We treat the case with three pump tones ($p=3$) in the same way as we did with two pumps above. By taking $\omega_{\Sigma}=\frac{\sum_{d=1}^{p}\omega_d}{p} = 2 \omega_r$, we have a  configuration of pump tones $\lbrace{-2\Delta,0,2\Delta\rbrace}$ with respect to $2 \omega_r$. We define our spectral modes around $\omega_r$ as $\lbrace(-2\Delta,-\Delta);(-\Delta,0);(0,\Delta);(\Delta,2\Delta)\rbrace$  $\widehat{=} \lbrace \tilde{a}_1(\omega); \tilde{a}_2(\omega); \tilde{a}_3(\omega); \tilde{a}_4(\omega) \rbrace$.
Individual pump tones at $\omega_1$ and $\omega_3$ create TMS states between neighboring spectral modes as in the 2 pump case above. However, the middle pump creates TMS correlations between two pairs of spectral modes: $\tilde{a}_1(\omega) \leftrightarrow \tilde{a}^\dagger_4(-\omega)$ and $\tilde{a}_2(\omega) \leftrightarrow \tilde{a}^\dagger_3(-\omega)$.  Rotation of the middle pump phase has an effect on both corresponding subspaces of the covariance matrix. Consequently, this pump configuration is quite suitable for producing square $\mathscr{\tilde{H}}$-graph states (see Fig. \ref{Nmodegraph}).

\begin{align}
    \mathbf{M}=\begin{bmatrix}
c & 0 & 0 & 0 & 0 & -\alpha & 0 & \pmb{-\alpha}\\
0 & c & 0 & 0 & -\alpha & 0 & \pmb{-\alpha} & 0\\
0 & 0 & c & 0 & 0 & \pmb{-\alpha}& 0 & -\alpha\\
0 & 0 & 0 & c & \pmb{-\alpha}& 0 & -\alpha & 0\\
0 & -\alpha& 0 & \pmb{-\alpha}& c & 0 & 0 & 0\\
-\alpha & 0 & \pmb{-\alpha} & 0 & 0 & c & 0 & 0\\
0 & \pmb{-\alpha}& 0 & -\alpha & 0 & 0 & c & 0\\
\pmb{-\alpha}& 0 & -\alpha & 0 & 0 & 0 & 0 & c\\
\end{bmatrix},
\label{intMatrix4modes}
\end{align}

The parametric interactions in the covariance matrix can be analysed in the same way as above, but now the number of phase differences influencing the beamsplitter correlations has increased. The system of linear equations ban be written as for three modes in Eq. (\ref{QLE3modes}), but we skip it and write down the interaction matrix $\mathbf{M}$ (Eq. (\ref{intMatrix4modes})), where all $c$ coefficient are equal since we have assumed $\Delta_r=\omega_r-\frac{\omega_{\Sigma}}{2}=0$. The signs of $\alpha$'s are governed by the choice of pump phases as $ \lbrace{ \alpha e^{\frac{i\pi}{2}}; \alpha e^{\frac{i\pi}{2}}; \alpha e^{\frac{i\pi}{2}}\rbrace}$. The correlations produced by the pump at $\omega_2=2\omega_r$ are indicated in bold. The special role of the central pump is seen because its correlations cover the whole ascending diagonal. 

The inverse matrix $\mathbf{M}^{-1}$ reveals the beamsplitter correlations between $\tilde{a}_1(\omega) \leftrightarrow \tilde{a}_3(\omega)$ and $\tilde{a}_2(\omega) \leftrightarrow \tilde{a}_4(\omega)$:

\begin{widetext}
\begin{eqnarray}
    \mathbf{M}^{-1}=\frac{1}{(c^2-4\alpha^2)}\cdot \nonumber \\ 
    \begin{bmatrix}
\begin{array}{cccccccc} c-\frac{2 \alpha ^2}{c} & 0 & \pmb{\frac{2 \alpha ^2}{c}} & 0 & 0 & \alpha & 0 & \alpha \\ 0 & c-\frac{2 \alpha ^2}{c} & 0 & \pmb{\frac{2 \alpha ^2}{c}} & \alpha & 0 & \alpha & 0 \\ \pmb{\frac{2 \alpha ^2}{c}} & 0 & c-\frac{2 \alpha ^2}{c} & 0 & 0 & \alpha & 0 & \alpha \\ 0 & \pmb{\frac{2 \alpha ^2}{c}}& 0 & c-\frac{2 \alpha ^2}{c} & \alpha & 0 & \alpha & 0 \\ 0 & \alpha & 0 & \alpha & c-\frac{2 \alpha ^2}{c} & 0 & \pmb{\frac{2 \alpha ^2}{c}} & 0 \\ \alpha & 0 & \alpha & 0 & 0 & c-\frac{2 \alpha ^2}{c} & 0 & \pmb{\frac{2 \alpha ^2}{c}} \\ 0 & \alpha & 0 & \alpha & \pmb{\frac{2 \alpha ^2}{c}} & 0 & c-\frac{2 \alpha ^2}{c} & 0 \\ \alpha & 0 & \alpha & 0 & 0 & \pmb{\frac{2 \alpha ^2}{c}}& 0 & c-\frac{2 \alpha ^2}{c} \\\end{array}
\end{bmatrix}
\label{interMatEq4modes}
\end{eqnarray}
\end{widetext}
The beamsplitter correlations are indicated in bold in this matrix $\mathbf{M}^{-1}$. We see that there are two sequences of pump transformations that yield BS correlations between modes $\tilde{a}_1(\omega) \leftrightarrow \tilde{a}_2(\omega)$ and $\tilde{a}_3(\omega) \leftrightarrow \tilde{a}_4(\omega)$. This agrees with the simple argument that indicates BS correlations to exist when two spectral bands are connected across squeezing action by two pumps with one joint frequency. Higher order correlations via three pumps exist also, but these are neglected in our analysis. Note that the number of beamsplitter correlations also coincides with the independent number of phase differences between the pumps. Connection of the cavity spectral mode correlations to  $\mathscr{\tilde{H}}$-graphs is illustrated in Fig. \ref{Nmodegraph}.

The beamsplitter correlations are prominent also in the covariance matrix $\mathbf{V}_\text{a}$  (see \ref{DEF_Va}):
\small
\begin{widetext}
\begin{eqnarray}
    \mathbf{V}_\text{a}= \frac{\kappa}{4(c^2-4\alpha^2)^2}\cdot \nonumber \\
    \resizebox{.9\linewidth}{!}{%
   $\begin{bmatrix}
     \begin{array}{cccccccc} -2 \alpha ^2+\frac{8 \alpha ^4}{c^2}+c^2 & 0 & 2 \alpha c & 0 & \pmb{ 6 \alpha ^2-\frac{8 \alpha ^4}{c^2} }& 0 & 2 \alpha c & 0 \\ 0 & -2 \alpha ^2+\frac{8 \alpha ^4}{c^2}+c^2 & 0 & -2 \alpha c & 0 & \pmb{ 6 \alpha ^2-\frac{8 \alpha ^4}{c^2} } & 0 & -2 \alpha c \\ 2 \alpha c & 0 & -2 \alpha ^2+\frac{8 \alpha ^4}{c^2}+c^2 & 0 & 2 \alpha c & 0 & \pmb{ 6 \alpha ^2-\frac{8 \alpha ^4}{c^2} } & 0 \\ 0 & -2 \alpha c & 0 & -2 \alpha ^2+\frac{8 \alpha ^4}{c^2}+c^2 & 0 & -2 \alpha c & 0 & \pmb{ 6 \alpha ^2-\frac{8 \alpha ^4}{c^2} } \\ \pmb{ 6 \alpha ^2-\frac{8 \alpha ^4}{c^2} } & 0 & 2 \alpha c & 0 & -2 \alpha ^2+\frac{8 \alpha ^4}{c^2}+c^2 & 0 & 2 \alpha c & 0 \\ 0 & \pmb{ 6 \alpha ^2-\frac{8 \alpha ^4}{c^2} } & 0 & -2 \alpha c & 0 & -2 \alpha ^2+\frac{8 \alpha ^4}{c^2}+c^2 & 0 & -2 \alpha c \\ 2 \alpha c & 0 & \pmb{ 6 \alpha ^2-\frac{8 \alpha ^4}{c^2} } & 0 & 2 \alpha c & 0 & -2 \alpha ^2+\frac{8 \alpha ^4}{c^2}+c^2 & 0 \\ 0 & -2 \alpha c & 0 & \pmb{ 6 \alpha ^2-\frac{8 \alpha ^4}{c^2} } & 0 & -2 \alpha c & 0 & -2 \alpha ^2+\frac{8 \alpha ^4}{c^2}+c^2 \\\end{array}
     \end{bmatrix}.$
     }
     \label{covarianceEq4modes}
\end{eqnarray}
\end{widetext}
\normalsize
 Bold font marks the beamsplitter correlations which display a different structure in comparison to Eq. (\ref{interMatEq4modes}) owing to the base change to quadratures ordered as $\left( \tilde{x}_1, \tilde{p}_1, \dots \tilde{x}_N, \tilde{p}_N \right)^{T}$. So the BS correlations are between quadratures of $\tilde{a}_1(\omega) \leftrightarrow \tilde{a}_3(\omega)$ and $\tilde{a}_2(\omega) \leftrightarrow \tilde{a}_4(\omega)$.

By choosing the phase of the first pump to be opposite to that of the second and the third $ \lbrace{ \alpha e^{\frac{-i\pi}{2}}; \alpha e^{\frac{i\pi}{2}}; \alpha e^{\frac{i\pi}{2}}\rbrace}$ we are able to flip the sign of one minor diagonal indicated by bold font in Eq. (\ref{intMatrix4modes_NOBS}). 
\vspace{-12pt}%
\begin{equation}
    \mathbf{M}=\begin{bmatrix}
c & 0 & 0 & 0 & 0 & \pmb{\alpha} & 0 & -\alpha\\
0 & c & 0 & 0 & \pmb{\alpha}  & 0 & -\alpha & 0\\
0 & 0 & c & 0 & 0 & -\alpha & 0 & -\alpha\\
0 & 0 & 0 & c & -\alpha & 0 & -\alpha & 0\\
0 & \pmb{\alpha}  & 0 & -\alpha & c & 0 & 0 & 0\\
\pmb{\alpha}  & 0 & -\alpha & 0 & 0 & c & 0 & 0\\
0 & -\alpha & 0 & -\alpha & 0 & 0 & c & 0\\
-\alpha & 0 & -\alpha & 0 & 0 & 0 & 0 & c\\
\end{bmatrix}
\label{intMatrix4modes_NOBS}
\end{equation}

Interestingly, this choice of phases leads to full cancellation of the beamsplitter correlation terms. This is seen in the structure of the inverse matrix:
\begin{eqnarray}
    \mathbf{M}^{-1}=\frac{1}{(c^2-2\alpha^2)}\cdot \nonumber\\ 
    \begin{bmatrix}
c & 0 & 0 & 0 & 0 & -\alpha & 0 & \alpha\\
0 & c & 0 & 0 & -\alpha & 0 & \alpha & 0\\
0 & 0 & c & 0 & 0 & \alpha & 0 & \alpha\\
0 & 0 & 0 & c & \alpha & 0 & \alpha & 0\\
0 & -\alpha & 0 & \alpha & c & 0 & 0 & 0\\
-\alpha & 0 & \alpha & 0 & 0 & c & 0 & 0\\
0 & \alpha & 0 & \alpha & 0 & 0 & c & 0\\
\alpha & 0 & \alpha & 0 & 0 & 0 & 0 & c\\
\end{bmatrix},
\label{interMatEq4modes_NOBS}
\end{eqnarray}

which does not have any elements proportional to $\alpha^2$. Without any beamsplitter correlations, Eq. (\ref{interMatEq4modes_NOBS})  indicates a clear connection to a square-lattice $\mathscr{\tilde{H}}$-graph. 

Finally, the corresponding covariance matrix $\mathbf{V}_\text{a}$ for cavity spectral modes is given by Eq. (\ref{covarianceEq4modes_NOBS}).
This structure for the covariance matrix is obtained when all the pump signals have an additional phase shift of $e^{\frac{i\pi}{2}}$. Such a choice of phases will result in a covariance matrix with "symmetric" structure as shown in experimental data in Figs. \ref{CovMat_exp}a and \ref{CovMat_exp}b. By controlling the phases of the pump tones separately, we can rotate and adjust certain subspaces of the $8 \times 8$ covariance matrix. In particular, the influence of the beamsplitter correlations can be eliminated from $\mathbf{V}_\text{a}$ in the four pump case. 

Regarding the quadripartite covariance matrix structures, the relative phase shift between the pump tones are not influenced by the cavity response in the limit of vanishing band widths or with the assumption of huge coupling rate and tiny internal dissipation loss rate. However, additional phase shifts will appear if these conditions are not met, which has to be taken into account in the generation of the desired entangled states.

In principle, it would be possible to evaluate the criteria for GME from the analytical expressions derived in this Appendix (see e.g. Eqs. (\ref{covarianceEq3modes9090phase}) and (\ref{covarianceEq4modes_NOBS})). However, 
we leave the conclusions about genuine entanglement both in the tripartite and quadripartite case for analysis based on numerical simulations where even the nonlinear terms can be taken into account. The nonlinear terms are of central importance when increasing the pump drive past the critical pumping amplitude.

\begin{widetext}
\begin{eqnarray}
    \mathbf{V}_\text{a}= \frac{\kappa}{4(c^2-2\alpha^2)^2}\cdot \nonumber \\
    \resizebox{0.8\linewidth}{!}{%
   $\begin{bmatrix}
     \begin{array}{cccccccc} c^2+2 \alpha^2 & 0 & -2 \alpha c & 0 & 0 & 0 & 2 \alpha c & 0 \\
     0 & c^2+2 \alpha^2 & 0 & 2 \alpha c & 0 & 0 & 0 & -2 \alpha c \\ 
     -2 \alpha c & 0 & c^2+2 \alpha^2 & 0 & 2 \alpha c  & 0 & 0 & 0  \\
     0 & 2 \alpha c & 0 & c^2+2 \alpha^2 & 0 & -2 \alpha c & 0 & 0 \\
     0 & 0 & 2 \alpha c & 0 & c^2+2 \alpha^2 & 0 & 2 \alpha c & 0 \\
     0 & 0 & 0 & -2 \alpha c & 0 & c^2+2 \alpha^2 & 0 & -2 \alpha c \\
     2 \alpha c & 0 & 0 & 0 & 2 \alpha c & 0 & c^2+2 \alpha^2 & 0 \\
     0 & -2 \alpha c & 0 & 0 & 0 & -2 \alpha c & 0 & c^2+2 \alpha^2\\\end{array}
     \end{bmatrix}.$
     }
     \label{covarianceEq4modes_NOBS}
\end{eqnarray}
\end{widetext}
\normalsize

\end{document}